\documentclass[11pt, a4paper]{article}
\usepackage{hyperref}
\usepackage{graphicx}
\usepackage[hscale=0.75,vscale=0.8]{geometry}
\usepackage{amsmath, amsthm, amssymb, mathrsfs}
\usepackage{color}
\usepackage{authblk}
\usepackage[dvipsnames]{xcolor}
\usepackage{enumerate}
\usepackage{mathtools}
\usepackage{todonotes}

\theoremstyle{definition}

\newcommand{\ii}{{\rm i}}
\newcommand{\ee}{{\rm e}}
\newcommand{\dd}{{\rm d}}

\newcommand{\vol}{{\rm vol}}

\begin{document}

\title{What happens once an accelerating observer has detected a Rindler particle?}
\author{Angel Garcia-Chung$^{1,2}$\thanks{{\tt alechung@xanum.uam.mx}}, Benito A. Ju\'arez-Aubry$^{3}$\thanks{{\tt benito.juarez@correo.nucleares.unam.mx}}, \\ and Daniel Sudarsky$^{3}$\thanks{\tt sudarsky@nucleares.unam.mx}}
\affil{$^{1}$Departamento de F\'isica, Universidad Aut\'onoma Metropolitana - Iztapalapa, \\
San Rafael Atlixco 186, Ciudad de M\'exico 09340, M\'exico.}
\affil{$^{2}$Tecnol\'ogico de Monterrey, Escuela de Ingenier\'ia y Ciencias, \\ 
Carr. al Lago de Guadalupe Km. 3.5, Estado de Mexico 52926, Mexico.}
\affil{$^{3}$Departamento de Gravitaci\'on y Teor\'ia de Campos, Instituto de Ciencias Nucleares, Universidad Nacional Aut\'onoma de M\'exico, A. Postal 70-543, Mexico City 045010, Mexico}

\date{\today}

\maketitle

\begin{abstract}
In a seminal paper, Unruh and Wald found that the detection of a right Rindler particle by a linearly uniformly accelerated detector coupled to a Klein-Gordon field in the Minkowski vacuum leads to the creation of a Minkowski particle from the inertial viewpoint. In this paper, we revisit the framework studied by Unruh and Wald, but now consider in addition what happens once the particle has been measured somewhere in the right Rindler wedge. From an orthodox point of view, the change in the field state induced by the measurement is non-local and occurs both in the left and right Rindler wedges. If one takes semiclassical gravity seriously in this context, this seems to open the possiblity for designing superluminal communication protocols between two spacelike separated observers confined to the right and left Rindler wedges respectively. We discuss the possible ways in which physics could prevent such measurement-induced, faster-than-light signaling protocols.
\end{abstract} 

\section{Introduction}

The Unruh effect posits that on a linearly uniformly accelerated trajectory the Minkwoski vacuum looks like a thermal state at a temperature proportional to the trajectory's proper acceleration \cite{Davies, Fulling, Unruh}. This is the Unruh temperature, given by $T_{\rm U} = a/(2 \pi)$ in natural units. The study of the Unruh effect continues to be of great interest in theoretical physics, see e.g. \cite{DeBievre:2006pys} for a study of the problem as return to equilibrium, \cite{Fewster:2016ewy} for the characterisation of thermalisation time for the Unruh effect, \cite{Salton:2014jaa} in the context of entanglement harvesting. The intimate relationship between the Unruh effect and Hawking radiation by black holes in equilibrium, which indeed originally motivated Unruh's work \cite{Unruh}, is by now well established (see e.g. \cite[Chap. 5]{WaldBook}), and can be seen very explicitly in two-dimensional situations \cite{Juarez-Aubry:2014jba}. \cite{Crispino, Hu:2012jr} include thorough discussions of the Unruh effect, including applications and structural properties. Very recently a concrete experimental proposal has been put forward in \cite{Gooding:2020scc} to detect for the first time the analogous Unruh temperature along uniformly accelerated circular motions. This is an analogous circular Unruh effect, see e.g. \cite{Biermann:2020bjh, Juarez-Aubry:2019gjw, Good:2020hav}.

Sitting at the heart of the Unruh effect is the fact that the Minkowski vacuum state restricted to, say, the right Rindler wedge of Minkowski spacetime, $|t| < x$, can be formally represented as thermal mixture of so-called \emph{Rindler particles} supported on the right Rindler wedge. These are nothing but particles defined in the Fulling-Rindler quantisation in flat spacetime, for which the notion of positive energy is defined with respect to Lorentz boosts, $b^a = a(x \, \partial_t^a + t\, \partial_x^a)$, which generate the natural notion of time evolution for linearly accelerated observers.

Following this observation, in 1984 Unruh and Wald wrote a seminal paper \cite{Unruh-Wald} where they clarified what occurs when a linearly uniformly accelerated observer detects a Rindler particle: From the point of view of an inertial observer in Minkowski spacetime, the absorption of a Rindler particle -- modelled as a two-level detector excitation -- corresponds to the emission of a Minkowski particle. The paper \cite{Unruh-Wald} is remarkable in that not only did it illustrate the relativity of the notion of a particle, detection and emission, but clarified the fact that working in terms of quantum fields (and taking the notion of particle to have a {\it contextual} and relative meaning) is fully consistent with the basic ideas underlying the equivalence principle.

Furthermore, \cite{Unruh-Wald} has served as the starting ground for further developments. For example, the study of \emph{bremsstrahlung} as seen from the point of view  of accelerated  observers \cite{Higuchi:1992td, Higuchi:1992we}, the analysis of the decay of accelerated protons, and  the finding that such  behavior approaches that of accelerated neutrons, as the mass scale characterizing that acceleration -- i.e.,  the   corresponding Unruh temperature --    increases, and  disappears  exponentially as  that   quantity  grows  beyond the  value of the proton-neutron mass gap \cite{Matsas-1, Matsas-2}.

In any case, there are three central issues that are addressed in \cite{Unruh-Wald}. The first one is to analyse the unitary evolution of the joint field-detector system when the field is initially in the Minkowski vacuum state and the two-level detector, initially switched off and prepared in the ground state, follows a linearly uniformly accelerated trajectory in the right Rindler wedge. This is done using perturbation theory in the interaction picture up to first order. (Second order contributions were further studied in \cite{Audretsch:1994gg}.) The second question is to see what the updated state of the field is,  assuming the detector has in fact detected a Rindler particle after some interaction time has elapsed, namely a one-particle state in the Minkowski folium, and to obtain the updated stress-energy tensor. It is found that, since the updated field state is a one-particle state, the energy of the field has increased upon detection. The third question addressed in \cite{Unruh-Wald} is whether detecting Rindler particles can be used as a mechanism for extracting an unbounded amount of energy from the field or to send a superluminal signal from the right to the left Rindler wedges. In both cases, the analysis leads to a negative answer.

The motivation of the present work is two-fold. First, we wish to revisit the three central questions discussed in \cite{Unruh-Wald}. Concerning the first one, we note that the calculation in \cite{Unruh-Wald} is performed by exploiting an analogy with the situation of a detector interacting with a field in a thermal state that describes a \emph{proper mixture} \cite{DEspagnat}, i.e., such that the actual state  of the system is pure, but there is a degree of ignorance as to what the state of the system actually is, which is encoded in weights accounting for a probability distribution of the possible (pure) states the system might be in. This results in a mixed state description of a pure state due to ignorance. On the other hand, the most natural description for the Minkowski vacuum from the point of view of an accelerated observer is that of a thermal state as an \emph{improper mixture} \cite{DEspagnat} (see footnote 2 for more details), as the left Rindler wedge degrees of freedom must be traced out, yielding a reduced mixed state. In sec. \ref{sec:Main} we will carry out the calculation from the improper-mixture viewpoint. While the results coincide mathematically, as they should, we think that this treatement is conceptually clearer. 

We then proceed to calculate the updated expectation value of the stress-energy tensor once the detector has clicked. We obtain expressions in both the right and the left Rindler wedge, adding to the result displayed in \cite{Unruh-Wald} for the right Rindler wedge, as we show in sec. \ref{sec:Main}. Concerning the third central question addressed in \cite{Unruh-Wald}, on the point of energy extraction, we agree with the no-go argument presented by Unruh and Wald: while the energy is \emph{not conserved} for a single measurement, it is  conserved on average for very many successive measurements. The discussion on superluminal communication ties in with the second motivation of this paper:

Here we shall raise the point that there is a potential issue after a single measurement has been carried out, if one is to trust the semiclassical approximation of quantum gravity ``before" and ``after" the measurement has been performed. The point will be that in semiclassical gravity the expectation value of the stress-energy tensor can be used to actually source geometry, see eq. \eqref{semi-simple} below. Thus, an abrupt change of this quantity might be  expected to be detectable by an experiment on the gravitational sector. Where and how this abrubt change occurs, i.e., where and how the state of the field can be seen as collapsing upon a measurement of the detector  is  most likely  to play a r\^ole on how to prevent this apparent paradox from occurring, but our current understanding of these questions is fuzzy -- hence the use of inverted commas around the words \emph{before} and \emph{after}. Thus, it seems to us that the resolution of this  puzzling   situation is likely    connected with the full resolution of  \emph{measurement problem} of quantum theory. We  will in fact   offer at the   end what we think is  a rather exhaustive list of possibilities for preventing  such   superluminal signals.

On this point we wish to emphasise that, while Unruh and Wald correctly point out in their discussion in \cite[Sec. IV]{Unruh-Wald} that the presence of a detector (switched on or otherwise) in the right Rindler wedge has no influence on the left Rindler wedge, their argument is based on the Heisenberg-picture observation that the effects of the detector can only affect the causal future of the coupling region between the detector and the field. In fact, this observation does not even depend on the details of the detector or the field observables, see e.g. \cite{Fewster:2018qbm} for a precise statement in some generality. The limitation of that argument is that it does not take into account the actual  measurement the detector, which is typically described as a projection onto the out-state in the interaction picture. This is a central difference between the above mentioned work and the  posture explored   in this paper.

At this point it is  worth mentioning that in the context of non-relativistic quantum mechanics there is a result known as the no-signaling theorem, which shows that entanglement between two separated systems cannot be used for  superluminal signaling. This result has, at this time, no  counterpart in QFT where how to deal with measurement processes, { is under development.} 
 Furthermore, in this case we will be considering the problem within the  semiclassical context for the treatment of gravitation. Moreover the no-signaling theorem involving joint measurements or manipulations made at one ``time'' on both components of the entangled system. The situation envisioned in this work concerns, as we will see, waiting arbitrarily large times for a detector to get excited and also waiting arbitrarily large times for the manifestation of an effect on the other side.

The organisation of this paper is as follows: In Sec. \ref{sec:Main} we describe the evolution of the field-detector system using a left and right doubled Fock space representation for the field, and we obtain the stress-energy tensor after a Rindler particle has been measured. In doing so, we do not make any assumptions on the details of the coupling of the detector and the field, in particular we do not assume a long-time limit for the interaction, other than assuming that the coupling is week, which allows us to conform ourselves with first-order effects in the coupling. We then discuss the non-conservation of energy upon measurements in Sec. \ref{sec:Energetic} in a simplified setting, for the sake of clarity. The possibility of faster than light signaling, its implications and potential paths for their  avoidance appear in Sec. \ref{sec:Superluminal}. Discussions and conclusions appear in Sec. \ref{sec:Conclusions}.

\section{What happens once an accelerating observer has detected a Rindler particle} \label{sec:Main}
    
Consider as in \cite{Unruh-Wald} a particle detector coupled to a Klein-Gordon field in Minkowski spacetime following a linearly uniformly accelerated trajectory with acceleration $a$ in the right Rindler wedge. In other words, the particle detector follows the integral curve generated by the boost $b^a = a(x \partial_t^a + t \partial_x^a)$. While currently the pointlike Unruh-DeWitt detector \cite{DeWitt, Louko:2007mu, Fewster:2016ewy} is the most prominent detector model  used in studies  about  the relativistic quantum information and QFT in curved spacetime literature, we shall model our detector as Unruh and Wald have in \cite{Unruh-Wald} to stay closer to their original treatment.

The detector is a two-level system with Hilbert space $\mathbb{C}^2$ spanned by energy eigenstates $|\uparrow\rangle$ and $|\downarrow\rangle$. The detector Hamiltonian is $\widehat H_{\rm D} := \Omega \widehat{A}^* \widehat{A}$, where $\widehat{A}$ and $\widehat{A}^*$ are raising and lowering operators, and $\Omega>0$ is the energy of the excited state, i.e., $\widehat H_{\rm D} |\uparrow\rangle = \Omega |\uparrow\rangle$ and $\widehat H_{\rm D} |\downarrow\rangle = 0$.

The coupled detector-field theory is described by the interaction Hamiltonian
\begin{align}
\widehat H = \widehat H_{\rm D} \otimes 1\!\!1 + 1\!\!1_{\rm D} \otimes \widehat H_\Phi + \widehat H_I, 
\end{align}
where $\widehat H_\Phi$ is the Klein-Gordon Hamiltonian and the interaction Hamiltonian is defined by
\begin{eqnarray}
\widehat{H}_I(\tau) = \epsilon(\tau) \int_\Sigma e^{ 2 a \xi} \dd\xi \, \dd y \dd z  \left[ \psi(\xi, y,z) \widehat{A}(\tau) + \overline{\psi}(\xi, y,z) \widehat{A}^*(\tau) \right] \otimes \widehat{\Phi}(\tau, \xi, y,z),
\end{eqnarray}
where $\widehat{\Phi}$ is the Klein-Gordon field, $\psi \in C_0^\infty(\Sigma)$ defines the profile of the spatial extension of the detector and $\epsilon \in C_0^\infty(\mathbb{R})$ is a switching function that controls the interaction between the detector and the Klein-Gordon field along the linearly uniformly accelerated trajectory of the detector. We shall assume that the interaction between the detector and the field is weaker than any other scale in the problem and that it takes place for sufficiently long times.

The coupling takes place in the right Rindler wedge, where the flat metric can be written in terms of Rindler coordinates
\begin{align}
t = \frac{1}{a} e^{a \xi} \sinh(a \tau), \qquad x = \frac{1}{a} e^{a \xi} \cosh(a \tau).
\end{align}

It takes the form
\begin{align}
\dd s^2 = -e^{2 a \xi} \left( \dd \tau^2 - \dd \xi^2 \right) + \dd y^2 + \dd z^2. \label{RRMetric} 
\end{align}

Furthermore, in the right Rindler wedge, the Klein-Gordon field can be written as
\begin{align}
\widehat \Phi(\tau, \xi, y, z) = \int_{\mathbb{R}^+\times \mathbb{R}^2} \dd^3 \kappa \left(v_{I \vec{\kappa}}(\tau, \xi, y, z) \widehat{a}_{{\rm R} \vec{\kappa}} + \overline{v_{I \vec{\kappa}}}(\tau, \xi, y, z) \widehat{a}_{{\rm R} \vec{\kappa}}^* \right), \label{PhiRight}
\end{align}
with $\vec{\kappa} = (\omega, \kappa_y,  \kappa_z)$, and where the right Rindler modes can be written in terms of the modified Bessel function of the second kind or MacDonald's function,
\begin{align}
v_{I \vec{\kappa}}(x) = \sqrt{\frac{ \sinh\left( \frac{\pi \omega}{a}\right) }{4 \pi^4 a}} K_{i \omega/a}\left[ \frac{\sqrt{\kappa^2_y + \kappa^2_y + m^2}}{a} e^{a \xi} \right] e^{-i \omega \tau + i \left( y \kappa_y + z \kappa_z\right)}  \label{Solg}
\end{align}
 and the formal sharp-momentum annihilation and creation operators are $\widehat{a}_{{\rm R} \vec{\kappa}} := \widehat{a}(\overline{v_{{\rm R} \vec{\kappa}}})$ and $\widehat{a}_{{\rm R} \vec{\kappa}}^* := \widehat{a}^*(v_{{\rm R} \vec{\kappa}})$ respectively. The annihilation operator annihilates the right Fulling-Rindler vacuum, $\Omega_{\rm R}$, while creation operators create Rindler particles.
 
A fully analogous description of the quantum theory holds in the left Rindler wedge. Introducing the left Rindler coordinates 
\begin{align}
t = \frac{1}{a} e^{a \tilde \xi} \sinh(a \tilde \tau), \qquad - x = \frac{1}{a} e^{a \tilde \xi} \cosh(a \tilde \tau)
\end{align}
the field in the left Rindler wedge takes the analogous form
\begin{align}
\widehat \Phi(\tilde \tau, \tilde \xi, y, z) = \int_{\mathbb{R}^+\times \mathbb{R}^2} \dd^3 \kappa \left(v_{II \vec{\kappa}}(\tilde \tau, \tilde \xi, y, z) \widehat{a}_{{\rm L} \vec{\kappa}} + \overline{v_{I \vec{\kappa}}}(\tilde \tau , \tilde \xi, y, z) \widehat{a}_{{\rm L} \vec{\kappa}}^* \right), \label{PhiLeft}
\end{align}
where the left Rindler modes have an identical form to the right modes \eqref{Solg} upon the replacement of $\tau$ and $\xi$ by $\tilde \tau$ and $\tilde \xi$. It is very well known that the Minkowski vacuum restricted to the (right or left) Rindler wedges looks like a thermal mixture of (right or left, resp.) Rindler particles. See appendix \ref{App:MinkoVac} for details.

In any case, in the case at hand we consider that the state of the system is prepared initially (before the switch-on of $\epsilon$) as the tensor product
\begin{align}
|s_{-\infty} \rangle = |\downarrow\rangle \otimes |\Omega_{\rm M} \rangle.
\end{align}

In the interaction picture, the late-time state of the system (after the switch-off of $\epsilon$) is given by
\begin{align}
|s_{\infty} \rangle & = T \left(\ee^{-\ii \int_\mathbb{R} \dd \tau \widehat H_I} \right) |s_{-\infty} \rangle = |s_{-\infty} \rangle + \left[- \ii \int_\mathbb{R} \dd \tau \widehat H_I + O(\epsilon^2) \right] |s_{-\infty} \rangle,
\label{s-infty}
\end{align}
under the assumption that the coupling is weak. To first order in perturbation theory, the late-time state of the system is
\begin{align}
|s_{\infty} \rangle = |\downarrow\rangle \otimes |\Omega_{\rm M} \rangle - i | \uparrow \rangle \otimes \int_{\mathbb{R}^+ \times \mathbb{R}^2} \!\!\!\!\!\!\!\!\!\! \dd^3 \kappa \int_{I} \dd \vol(x) \zeta(x) \left( v_{I \vec \kappa}(x) \hat a_{{\rm R} \vec \kappa} + \overline{v_{I \vec \kappa}(x)} \hat a^*_{{\rm R} \vec \kappa}  \right) |\Omega_{\rm M}\rangle ,  \label{LateState}
\end{align}
where the volume element in the right Rindler wedge is locally $\dd \vol(x) = e^{2 a \xi} \dd \tau \dd \xi \dd y \dd z$ and 
\begin{align}
\zeta(x) := e^{i \Omega \tau} \varepsilon(\tau) \overline{\psi(\xi,y,z)}.
\end{align}

In \cite{Unruh-Wald} the assumption has been made that $\varepsilon(\tau)$ is nearly constant, physically representing a long interaction between the detector and the field, such that switching effects are negligible. In this case, the $\tau$ integral can be performed directly and one obtains a factor proportional to $\delta(\Omega-\omega)$ that indicates that only the modes $v_{I \vec \kappa}$ with frequency highly localised around $\Omega$ will contribute to first order in eq. \eqref{s-infty}. This makes perfect sense, after a long interaction time the only modes that get excited from the vacuum state are those whose energy coincides with the frequency gap of the detector.

Let us however digress at this point and not make this approximation. The reason is that in our case we are interested in post-measurement effects for measurements carried out after a finite time of interaction between the field and the detector. On the assumption that the detector clicks, the updated state for the field becomes
\begin{align}
|f\rangle = -i \mathcal{N} \int_{\mathbb{R}^+ \times \mathbb{R}^2} \!\!\!\!\!\!\!\!\!\! \dd^3 \kappa \int_{I} \dd \vol(x) \zeta(x) \left( v_{I \vec \kappa}(x) \hat a_{{\rm R} \vec \kappa} + \overline{v_{I \vec \kappa}(x)} \hat a^*_{{\rm R} \vec \kappa}  \right) |\Omega_{\rm M}\rangle,
\end{align} 
where the normalisation $\mathcal{N}$ is given by
\begin{align}
\mathcal{N} = \left( \int_{\mathbb{R}^+ \times \mathbb{R}^2} \!\!\!\!\!\!\!\!\!\! \dd^3 \kappa  \int_{I} \dd \vol(x)  \int_{I} \dd \vol(x') \overline{\zeta(x)} \zeta(x') \left( \frac{v_{I \vec \kappa}(x)\overline{v_{I \vec \kappa'}(x')}}{1- e^{-2 \pi \omega/a}} + \frac{\overline{v_{I \vec \kappa}(x)}  v_{I \vec \kappa'	}(x')}{e^{2 \pi \omega/a}-1} \right) \right)^{-1/2}
\label{Normalisation}
\end{align}
as can be seen in appendix \ref{App:Normalisation}.

We are interesting in the change in the expectation value of the stress-energy tensor of the field in the updated state, i.e., we are interested in
\begin{align}
\Delta T_{ab} := \langle f | \hat T_{ab}(x) f \rangle - \langle \Omega_{\rm M} | \hat T_{ab} \Omega_{\rm M} \rangle
\end{align}
in the right and left Rindler wedges. (Note that imposing $\langle \Omega_{\rm M} | \hat T_{ab} \Omega_{\rm M} \rangle = 0$,  we have that $\Delta T_{ab} = \langle f | \hat T_{ab}(x) f \rangle$.) To this end, if we use a point-splitting prescription for renormalising the stress-energy tensor the object of interest is to obtain the two-point function in the left and right Rindler wedges. It follows from the calculations in appendices \ref{Sec:RightUpdate} and \ref{Sec:LeftUpdate} that the two-point function in the updated state takes the form
\begin{align}
\langle f | \hat \Phi(x) \hat \Phi(x') f \rangle & = \langle \Omega_{\rm M} | \hat \Phi(x) \hat \Phi(x') \Omega_{\rm M} \rangle+ \Delta_{\rm R}(x,x') \text{ in the right Rindler wedge and,} \\
\langle f | \hat \Phi(x) \hat \Phi(x') f \rangle & = \langle \Omega_{\rm M} | \hat \Phi(x) \hat \Phi(x') \Omega_{\rm M} \rangle+ \Delta_{\rm L}(x,x') \text{ in the left Rindler wedge}.
\end{align}

Here, $\Delta_{\rm R}$ and $\Delta_{\rm L}$ are real, smooth, symmetric bi-functions given by
\begin{align}
\Delta_{\rm R}(x,x') & = \mathcal{N}^{2} \int_{\mathbb{R}^+ \times \mathbb{R}^2} \!\!\!\!\!\!\!\!\!\! \dd^3 \kappa   \int_{\mathbb{R}^+ \times \mathbb{R}^2} \!\!\!\!\!\!\!\!\!\! \dd^3 p  \int_{I} \dd \vol(y)  \int_{I} \dd \vol(y') \frac{\overline{\zeta(y)} \zeta(y')}{(1-e^{-2\pi \omega_p/a}) (1-e^{-2\pi \omega_\kappa/a})} \nonumber \\
& \times \left( v_{I \vec p}(y) v_{I \vec{\kappa}}(x) \overline{v_{I \vec{p}}(x')} \overline{v_{I \vec \kappa}(y')} + \overline{v_{I \vec \kappa}(y)} v_{I \vec{\kappa}}(x) v_{I \vec{p}}(x') \overline{v_{I \vec p}(y')}e^{-2 \pi \omega_\kappa} \right. \nonumber \\
& \left. + \overline{v_{I \vec p}(y)} v_{I \vec{\kappa}}(x) v_{I \vec{p}}(x') \overline{v_{I \vec \kappa }(y')}e^{-2 \pi \omega_p} + \overline{v_{I \vec \kappa}(y)} v_{I \vec{\kappa}}(x) \overline{v_{I \vec{p}}(x')} v_{I \vec p}(y')e^{-2 \pi \omega_{\kappa}} e^{-2 \pi \omega_{p}}\right) + {\rm c.c. }, \label{DeltaR} \\
\Delta_{\rm L}(x,x') & = \mathcal{N}^{2} \int_{\mathbb{R}^+ \times \mathbb{R}^2} \!\!\!\!\!\!\!\!\!\! \dd^3 \kappa  \int_{\mathbb{R}^+ \times \mathbb{R}^2} \!\!\!\!\!\!\!\!\!\! \dd^3 p  \int_{I} \dd \vol(y)  \int_{I} \dd \vol(y') \overline{\zeta(y)} \zeta(y') \frac{e^{- \pi \omega_p/a} e^{- \pi \omega_\kappa/a} }{(1-e^{-2 \pi \omega_p/a})(1-e^{-2 \pi \omega_\kappa/a})} \nonumber \\
& \times \left( v_{I \vec p}(y) v_{II \vec{\kappa}}(x) \tilde v_{II \vec{p}}(x') \tilde v_{I \vec \kappa}(y') + v_{I \vec p}(y) \tilde v_{II \vec{p}}(x) v_{II \vec{\kappa}}(x') \tilde v_{I \vec \kappa	}(y') \right. \nonumber \\
& \left. + v_{I \vec p}(y) \overline{v_{II \vec{\kappa}}(x)} \tilde v_{II \vec{p}}(x') \overline{\tilde v_{I \vec \kappa}(y')} + \overline{v_{I \vec p}(y)} \overline{\tilde v_{II \vec{p}}(x)} v_{II \vec{\kappa}}(x') \tilde v_{I \vec \kappa	}(y')
 \right) + {\rm c.c. }, \label{DeltaL}
\end{align}  
where in eq. \eqref{DeltaL} the tilde on the modes denotes a parity operator in the orthogonal direction to the Rindler wedges, i.e., $\tilde v_{I \vec \kappa	} :=  v_{I (\omega_\kappa, -\kappa_\perp)}$ and similarly for $\tilde v_{II \vec \kappa	}$.

Let us now compute the expectation value of the stress-energy tensor in the state $| f \rangle$. Note that a difference with the calculation in \cite{Unruh-Wald} is that we do not treat the state of the system as an improper thermal mixture in the right Rindler wedge, but rather as a pure state defined in the right and left Rindler wedges, which allows us to obtain the renormalised expectation value of the stress energy tensor for points in the right and left Rindler wedges.

Here we use the terminology introduced by d'Espagnat  \cite{DEspagnat} to clarify that the  same  mathematical object, a density matrix, can be used to represent  two  very  different  physical situations:  1) The  case in   which  one  is interested  in an ensemble  of identical  quantum systems,  each one  of which is  in one  pure   and  definite quantum state    among a  list   possible   such  states $ \{   |i \rangle \}$   and where  the fraction  of  such states in the ensemble  is given by a certain  classical distribution function  $f (i)$, and 2) the case  in which  a    system of interest  $S$  is  a   subsystem of a larger  system $ S+E$ with the latter in a  given  pure quantum state,  but with our  interest focused just on $S$  which  can therefore  be characterised in  terms of the reduced  density matrix  obtained  after tracing over  the  degrees of freedom  of $E$.   For the  first case  one  reserves the name  ``proper mixture" and    says the  density  matrix   represents   such  proper mixture,  and for the second case   one reserves the name  ``improper  mixture", and  equally  indicates the  density matrix is to be understood as  representing the improper    mixture.     We note that another  situation one  might  want  to  consider,  is  one  in which  one   is dealing with a   single  quantum  system $S$ which is in  a pure state,  which however is not  completely known, and for which  one  has  information  about    the   classical probability  $p (i)$ of the system   being in  each  one   of  the quantum states.    For such situation one   can often  use the characterisation  provided  by   case 1)  by  considering  a   corresponding imaginary  ensemble,     in  which the  fraction is   arranged to  match the   given probability   i.e.,  $  f(i) = p(i)$.   That   situation one also   talks  by extension about a    proper mixture,  even though the state of the system  is   pure, and  thus its  ``properness"  (or the fact that we  do not express  the state  as a pure one)  is  just a result of our ignorance.   Finally, as is usual,  a    density matrix  is    characterised  as thermal  if  its  representation in the energy basis  has  the  standard thermal  weights.   Thus   a  thermal density  matrix  can be  proper or improper.

\section{The change in the stress-energy tensor}
\label{sec:ChangeTab}

In order to compute the stress-energy tensor in the updated state, $\langle f | \hat T_{ab}(x) f \rangle$, in the right/left wedge we apply the point-splitting operator
\begin{align}
\mathcal{T}_{ab} := g_b{}^{b'} \nabla_a \nabla_{b'} - \frac{1}{2} g_{ab} g^{cd'} \nabla_c \nabla_{d'} - \frac{1}{2} g_{ab} m^2,
\end{align}
where $g_a{}^{a'}$ is the parallel-transport propagator, to $\Delta_{{\rm R/L}}(x,x')$ given by eq. \eqref{DeltaSimp}, and then take the coincidence limit as
\begin{align}
\langle f | \hat T_{ab}(x) f \rangle = \lim_{x'\to x} \mathcal{T}_{ab} \Delta_{\rm R/L}(x,x').
\label{Tcoinc}
\end{align}

It is useful to locally express the parallel-transport components in Rindler coordinates by using the formula $g_\mu{}^{\mu'}(x,x') = e_\mu^I(x) e^{\mu'}_I(x')$ in terms of the soldering form and its inverse. We then have that in the right Rindler wedge the parallel-transport propagator has components $g_\eta{}^{\eta'}(x,x') = e^{a \xi} e^{-a \xi'}$, $g_\xi{}^{\xi'}(x,x') = e^{a \xi} e^{-a \xi'}$, $g_y{}^{y'} = 1$ and $g_z{}^{z'} = 1$, with all other components vanishing.

We are chiefly concerned with changes in the left Rindler wedge, which is causally disconnected from the detector that clicks. Inserting \eqref{DeltaL} into eq. \eqref{Tcoinc} we have that the in the left Rindler wedge stress-energy tensor in the updated state takes a diagonal form and the form it takes can be read directly from \eqref{Tcoinc}. For instance, for the energy density of a massless field we have in the left Rindler wedge
\begin{align}
& \langle f | \hat T_{\eta \eta}(x) f \rangle  = \Re \, \, \mathcal{N}^{2} \int_{\mathbb{R}^+ \times \mathbb{R}^2} \!\!\!\!\!\!\!\!\!\! \dd^3 \kappa  \int_{\mathbb{R}^+ \times \mathbb{R}^2} \!\!\!\!\!\!\!\!\!\! \dd^3 p  \int_{I} \dd \vol(y)  \int_{I} \dd \vol(y') \overline{\zeta(y)} \zeta(y') \frac{e^{- \pi \omega_p/a} e^{- \pi \omega_\kappa/a} }{(1-e^{-2 \pi \omega_p/a})(1-e^{-2 \pi \omega_\kappa/a})} \nonumber \\
& \times \left( v_{I \vec p}(y) \partial_\eta v_{II \vec{\kappa}}(x) \partial_\eta \tilde v_{II \vec{p}}(x) \tilde v_{I \vec \kappa}(y') +  e^{2 a \xi} \sum_{i = 1}^3 g^{ii} v_{I \vec p}(y) \partial_i v_{II \vec{\kappa}}(x) \partial_i\tilde v_{II \vec{p}}(x) \tilde v_{I \vec \kappa}(y')  \right. \nonumber \\
& + v_{I \vec p}(y) \partial_\eta \tilde v_{II \vec{p}}(x) \partial_\eta v_{II \vec{\kappa}}(x) \tilde v_{I \vec \kappa	}(y') + e^{2 a \xi} \sum_{i}^3 g^{ii} v_{I \vec p}(y) \partial_i \tilde v_{II \vec{p}}(x) \partial_i v_{II \vec{\kappa}}(x) \tilde v_{I \vec \kappa	}(y')  \nonumber \\
&  + v_{I \vec p}(y) \partial_\eta \overline{v_{II \vec{\kappa}}(x)} \partial_\eta \tilde v_{II \vec{p}}(x) \overline{\tilde v_{I \vec \kappa}(y')} + e^{2 a \xi} \sum_{i}^3 g^{ii} v_{I \vec p}(y) \partial_i \overline{v_{II \vec{\kappa}}(x)} \partial_i \tilde v_{II \vec{p}}(x) \overline{\tilde v_{I \vec \kappa}(y')} \nonumber \\
& \left. + \overline{v_{I \vec p}(y)} \partial_\eta \overline{\tilde v_{II \vec{p}}(x)} \partial_\eta v_{II \vec{\kappa}}(x) \tilde v_{I \vec \kappa	}(y') + e^{2 a \xi} \sum_{i}^3 g^{ii} \overline{v_{I \vec p}(y)} \partial_i \overline{\tilde v_{II \vec{p}}(x)} \partial_i v_{II \vec{\kappa}}(x) \tilde v_{I \vec \kappa	}(y')
 \right), \label{EnergyLeft}
\end{align}
where $\Re$ denotes the real part and with $g^{11} = e^{-2 a \xi}$ and  $g^{22} = g^{33} = 1$. In the massive case, one adds the term
\begin{align}
\frac{1}{2} m^2 e^{2 a \xi } \Delta_{\rm L}(x,x)
\label{MassTermEnergy}
\end{align}
to the right-hand side of eq. \eqref{EnergyLeft}.

Similar expressions can be obtained for the components $\langle f | \hat T_{\xi \xi}(x) f \rangle$, $\langle f | \hat T_{yy}(x) f \rangle$ and $\langle f | \hat T_{zz}(x) f \rangle$ in the left Rindler wedge, and for the four non-vanishing components in the right Rindler wedge.

Instead of spelling out in detail all of the remaining components, we point out that eq. \eqref{EnergyLeft} suffices to a central point of the paper, which is that a measurement that occurs in the right Rindler wedge has non-trivial effects on the causally-disconnected left Rindler wedge. It is natural to ask ``when" or ``where" in spacetime the state collapses after a detector has detected a Rindler particle. This is far from obvious, but a natural assumption seems to be that the state collapses along a Cauchy surface of spacetime (see \cite{Juarez-Aubry:2017ery}) intersecting the ``detection event" on the right Rindler wedge and extending into the left Rindler wedge. 

We should emphasise that regardless of ``how big" this change might be, it is a principled statement that by including state collapses with semiclassical gravity has produced an abrubt change in the a region causally disconnected from where the measurement took place, in this case by means of a detector click.

One can see by direct inspection of \eqref{EnergyLeft} and \eqref{MassTermEnergy} that for high accelerations there exists spacetime regions in the left Rindler wedge where the expectation value of the energy density becomes large. For example, for sufficiently small $\tilde \xi$, say $\tilde \xi \sim 1/a$, the Rindler modes do not exhibit the large-argument suppression of the MacDonald function, but the integrand factors $e^{-\pi \omega / a}/(1- e^{-2 \pi \omega/a})$ exhibit the behaviour
\begin{align}
\frac{e^{-\pi \omega / a}}{1- e^{-2 \pi \omega/a}} = \frac{1}{2\sinh (\pi \omega/a)} = \frac{a}{2 \omega}+ O(\omega/a).
\end{align}

This observation is consistent with what one would expect in the long interaction time limit case presented in \cite{Unruh-Wald}, as can be seen from eq. (3.29) in that paper in the small $\beta = 2\pi/a$ regime.

\section{Energetic considerations}
\label{sec:Energetic}

In this section we revisit  some  of the energetic considerations discussed in \cite{Unruh-Wald} but focusing on the various possible {\it individual outcomes} of the ``detection attempts", rather than on the ensemble averages of detector measurement outcomes, which are the quantities   considered in most of the discussion of said reference.

Following  \cite{Unruh-Wald} and to simplify the discussion we will consider   the case of  {(ensembles of)} harmonic oscillators,    and  entangled  pairs  of harmonic oscillators  instead of quantum fields. There is  no loss of conceptual clarity in doing so, but the treatment is mathematically less involved. 

Consider a harmonic oscillator with energy eigenstates  $\lbrace |n\rangle \rbrace $ with renormalised energies  $n\epsilon$ (i.e., we removed the   zero point or ground state energy for simplicity of analysis), and a detector with two states $ | \downarrow \rangle$ and $|\uparrow\rangle$  with energy levels  $0$  and $\epsilon$ respectively.
Let us assume that the initial state of the combined system is: 

       \begin{equation}
\label{state-1 }
  |\Psi\rangle = N \left( |0\rangle \otimes  |\downarrow\rangle   + \frac{1}{\sqrt{\alpha}} |n\rangle \otimes  |\uparrow\rangle  \right),
\end{equation} 
with $ N^2 = 1/(1+\alpha^{-1})$. The expectation value of the energy is $\langle E \rangle_{\Psi}=\frac{n \epsilon }{1+\alpha} $.

If an observer finds the detector in the unexcited state $ | \downarrow\rangle$, the value of the energy becomes $\langle E \rangle_{\rm unex} = 0$, and the probability for this  is ${\rm P}_{\rm unex}  =  N^2$. On the other hand  if they find the  detector in the excited state, the energy becomes $\langle E \rangle_{\rm ex} = n  \epsilon$ and the probability for this is ${\rm P}_{\rm ex}  =  N^2/ \alpha$. The average result is of course $\langle E \rangle  =\frac{n \epsilon}{1+\alpha}  $, which is the same as $\langle E \rangle_{\Psi}$.  We note however that in each specific case (i.e., for each specific outcome of the observation) the actual value of the energy differs from $\langle E \rangle_{\Psi}$, the expectation value of the initial state of the combined system. This is of course not surprising, given that the state $|\Psi \rangle$ is not an eigenstate of the total energy operator.

 As  has  been argued in \cite{Tim-Elias}, the relevant issue regarding energy conservation is not its conservation ``on average", but its conservation on each  single individual instance of any experiment. In this sense the possibility of preparing a state such as $|\Psi\rangle$ and to make the  observations described above already set serious doubts about the general validity of anything like  a law of ``energy conservation" in the quantum setting.

\subsection{Proper mixture}

Consider now the case of a system in thermal equilibrium at temperature $T$. We take the system in question  once  again  to  be   a simple  harmonic oscillator.  This  corresponds, as  is traditionally treated on statistical  mechanic textbooks,  to  an ensemble (a canonical  ensemble) of identical particles and might be  described  by the  proper mixture:
\begin{equation} \label{density-matrix-1}
  \rho = N  \Sigma_{n=0}^{\infty }  e^{-\beta n \epsilon } |n \rangle \langle n |,
\end{equation}
\noindent where $ N = 1-e^{-\beta \epsilon} $  is  a normalisation constant  ensuring ${\rm Tr} \rho =1 $.  The  mean energy  is then $\langle  E \rangle_{T}= \frac{\epsilon \, e^{-\beta \epsilon}}{1-e^{-\beta \epsilon}}$. Let us now  consider a two  level detector  (as  in the  previous discussion) initially in the un-excited state (and vanishing  energy) and make it interact with our thermal ensemble (for this purpose  we  in  fact  consider  an  ensemble of identical  detectors). The  initial  situation will thus  be  described  by the density matrix:
\begin{align}   \label{density-matrix-2 }
\rho\otimes |\downarrow \rangle \langle \downarrow | = \left[ \left( 1-e^{-\beta \epsilon}\right)  \Sigma_{n=0}^{\infty }  e^{-\beta n \epsilon } |n \rangle \langle n | \right] \otimes  |\downarrow \rangle \langle \downarrow | .
\end{align}

After   letting the  system  interact   for a suitably  long  time  we   will find  a result analogous to  that encountered  in  the     equation (3.25)   of \cite{Unruh-Wald}. That is:
\begin{align} \label{density-matrix-3}
  \rho_{\rm late} =  N_{\rm late} \; \Sigma_{n=0}^{\infty }  e^{-\beta n\epsilon}  \left[ |n\rangle \otimes | \downarrow \rangle   - i \gamma \sqrt{n} |n-1\rangle \otimes |\uparrow \rangle  + \dots ][
  \langle n | \otimes \langle \downarrow |  +  i \gamma\sqrt{n} \langle n-1| \otimes \langle \uparrow| + \dots \right]
\end{align}
 where $\gamma$ is  a small parameter  representing the  strength and time  duration  of the interaction and $  N_{\rm late}  $ is  a normalisation constant  ensuring ${\rm Tr} \, \rho_{\rm late} =1 $.

\subsection{Improper mixture}

Consider now the  situation  in which  we are informed that the  detector  is  excited. To do  this  we  simply compute a partial trace  after applying the  corresponding  projector   $  |\uparrow \rangle \langle \uparrow|$. The   resulting density matrix up, to first order in the expansion is
\begin{equation}  \label{density-matrix-4}
  \rho_{\rm post} = N_{\rm post}  \Sigma_{n=1}^{\infty }  e^{-\beta n\epsilon} n  |n-1\rangle  \langle n-1|,
\end{equation}
\noindent where $ N_{\rm post}=  \left( \frac{ e^{-\beta \epsilon}}{(1-e^{-\beta \epsilon})^2} \right)^{-1}$ is  still another normalisation constant  ensuring ${\rm Tr} \, \rho_{\rm post} =1 $.  

 We note that the  absence of the  first  term  ought   be considered  to be the   result of   both the  measurement of the excitation level  of the  exited   detector  and   the  post selection  within the  ensemble  in  which the   elements  containing   the  un-excited   detector state  were removed. In other words,  we have  that in going from  eq. \ref{density-matrix-3} to eq. \ref{density-matrix-4}  actually modifies  the   set of systems  composing  the ensemble  itself.

      The  next  observation is that the resulting  ensemble, represented  by the  density matrix \ref {density-matrix-4}  is not longer thermal, which in turn  implies that the expectation  value of the energy 
\begin{equation}  \label{energy-post}
{\rm  Tr} ( \hat H \rho_{\rm post}) = N_{\rm post}  \Sigma_{m=0}^{\infty }  e^{-\beta (m+1) \epsilon} (m+1)m \epsilon =  \frac {2 \epsilon e^{-\beta \epsilon}}{(1-e^{-\beta \epsilon})}.
\end{equation}
differs  slightly from   that of a  truly thermal ensemble ($\langle  E \rangle_{T}= \frac{\epsilon \, e^{-\beta \epsilon}}{1-e^{-\beta \epsilon}}$).
We note  again that even  after adding the  energy  of excitation of the  detector $ \epsilon$ the expectation value of the  energy has  changed  as  the  result of the measurement (and  post-selection).

       Finally  let us   consider the case of  a  single pair of  entangled  harmonic  oscillators  with ``thermal  weights". That is the case of  a pure  state, which  upon  consideration of the reduced  density matrices  for  each  oscillator would  result  in a thermal density matrix,  but of course  one of  an improper nature.     Let us  add a  detector arranged to interact just with  the harmonic    oscillator  (II) and which is   initially prepared in the  unexcited  state, so the state of the complete system is:
   \begin{equation}
\label{entangledstate-1}
 {|\Psi \rangle}^{(T)}_{\rm initial} =  A_{\rm initial}  \Sigma_{n=0}^{\infty }  e^{-\beta n\epsilon} |n \rangle^{(I)} \otimes  |n\rangle^{(II)} \otimes | \downarrow\rangle,
\end{equation} 
where $ A_{\rm initial} =   \sqrt{1-e^{-2 \beta \epsilon}} $. After  letting the  system  interact   for a suitably  long  time,  the state of the  system  will be : 
\begin{equation}
    \label{entangledstate-2}
  |\Psi \rangle^{(T)}_{\rm late} = A_{\rm late} \Sigma_{n=0}^{\infty}  e^{-\beta n\epsilon} |n \rangle^{(I)} \otimes \left[ |n \rangle^{(II)}  \otimes |\downarrow \rangle  - i  \gamma\sqrt{n} |n-1 \rangle^{(II)} \otimes |\uparrow \rangle  + \dots \right].
\end{equation}
If we  project on the subspace  corresponding to the excited  detector (ignoring the irrelevant factor $\ii$ and ensuring that the final state is  normalised) we find: 
 \begin{equation} \label{entangledstate-3}
|\Psi \rangle^{(T)}_{\rm pos-tsel} = A_{\rm post-sel}  \Sigma_{n=0}^{\infty }  e^{-\beta n\epsilon} |n \rangle^{(I)} \otimes \left[  \sqrt{n} |n-1\rangle^{(II)} + \dots \right],
\end{equation}
\noindent  where in this case $ A_{\rm post-sel} \approx  \left( \frac{ e^{-2 \beta \epsilon}}{(1-e^{- 2 \beta \epsilon})^2} \right)^{- 1/2}$. 

  Now  the  expectation values of the energy of  the harmonic  oscillator $I$  is  $ \langle E \rangle_I =     \frac { \epsilon  (1+ e^{- 2\beta \epsilon}) }{(1-e^{-2 \beta \epsilon})} $, which is different from that of  a purely thermal state,  while the expectation value of the harmonic  oscillator $II$   is  $ \langle E \rangle_{II} =   \frac {2  \epsilon  e^{- 2 \beta \epsilon}}{(1-e^{- 2\beta \epsilon})} $. Note that if we  add the  energy  of the excited detector,  $\epsilon$,  we have that $ \langle E \rangle_{II} + \epsilon   =  \langle E \rangle_I  $.  On the  other hand  the  expectation  energies  of  both cases  have  definitely   changed  as a result of the  ``projection".

 At this point  we might find  nothing seriously  puzzling   because  we have   been dealing with  systems  that   are  not  initially in  energy  eigenstates. The  change in the case of the harmonic   oscillator II in the  above example,   is of course  a bit puzzling  due to the  fact that   this  oscillator has not been  made to directly interact  with  a  detector to bring about the ``projection", however this   is  nothing  more  than the  usual  change occurring as a  result of quantum  entanglement with a second  system which has  been subjected to a measurement. 
 
 It is  worth noting that,  in  dealing with mixed states it is  only  when  the  expectation   of the  energy momentum  is    extracted   from an improper mixture (and thus, indirectly from a pure  state)   that it make  sense to use it in semiclassical gravity.  If this is done  with a  proper mixture what  we  would   obtain is  indeed  an average   value of that quantity  over  some  ensemble (such as  in the  case of the  many   realizations involved  in  stochastic  gravity) and then  self consistency will be  in  doubt\footnote{ In fact we  do not know  what it  means to  make  averages  over collections of space-times  and the non-linearity  of GR clearly cast  serious doubts that say  Einstein's  equations  would  be  preserved  under any kind  of  averaging  one  might   want to consider.}.       

 The situation  considered  in Sec. 4  is,  however  a bit more troublesome as it  seems to have the  potential  for a   serious violation of  our fundamental  ideas  about  relativity, in particular, the  potential  to offer  a path  for  superluminal communication.

 \section{Faster than light signaling?}      
\label{sec:Superluminal}
           
   It is by now  widely agreed that our world contains non-local features. This is reflected for instance in the  violations of Bell's  inequalities\cite{Bell}, which have been experimentally confirmed by now in several experiments \cite{Bell experiments, Bell experiments-2, Bell experiments-3, Bell experiments-4, Bell experiments-5}. While even simpler theoretical settings, such as the GHZ construction  \cite{GHZ} (which  have not been experimentally realised due to technical difficulties) are expected  to provide further  and   even more transparent evidence for non-locality. See   for instance   the discussion  about the  GHZ  scheme in \cite{Maudlin-Book}.

Nevertheless, there are a number of arguments and general widespread conviction that such non-locality cannot be exploited to communicate superluminally. Indeed, the non-locality present in the situation examined by Bell does not allow for superluminal communication. Faster-than-light communication  would force us to revise the very foundations of  special relativity -- and of physics as a whole.  The widespread posture in the physics community is that somehow nature contains features that prevents the Bell type non-locality, and in general the nonlocal nature of the quantum state,  (which by the way cannot be taken as being of pure epistemic nature as per the PRB theorem \cite{PRB}), from being used for superluminal signaling. In any case, ``textbook'' quantum mechanics and its alternatives seeking to resolve the measurement problem do not seem to offer paths that allow for faster-than-light communication. This is captured by the set of results known as the no-signaling theorem.  

Things become more complicated in the context of quantum field theory, as the question of what quantities (represented by, say, self adjoint operators), can be measured or not is an issue with not complete and generally accepted answer, see \cite{Sorkin Imposible} and a possible resolution \cite{Bostelmann:2020unl}.  The  example we study here  involves  yet  another  aspect that  seems problematic.

The change that we have noted in the expectation value of the stress-energy tensor in the left Rindler wedge (region II) due to measurements in the right Rindler wedge (region I) -- if detectable -- however seems to provide a path for superluminal communication, which we describe in the following \emph{gedankenexperiment}:

{\bf \emph{Gedankenexperiment}:} Suppose a linearly uniformly accelerated observer in the right Rindler wedge, say Alice, is equipped with a highly-efficient detector for which it is highly probable to detect a Rindler particle and non-detection is negligible. This enables Alice to change the expectation value of the stress-energy tensor in the left Rindler wedge by turning her detector on or to decide not to do it by not switching on her detector. A causally-disconnected observer, say Bob, in the left Rindler wedge that can probe, either the state of the field  or  more specifically, the expectation value of  stress-energy tensor (for example by probing the gravitational field with a torsion balance),  would then be able to infer whether Alice has or has not turned on her detector in the right Rindler wedge. This seems to be a path for achieving  superluminal communication between Alice and Bob.

A simple specific protocol that realises the above {\it Gedankenexperiment} can be thought of as follows.

 The  simplest  specific  protocol  for  superluminal  communication  is the   following.  Alice and Bob  are given  instructions  that Alice could  send  to Bob  a  signal  (or  instance that  she has decided on   something   and the  answer is ``yes")   by  turning on  her  detector.  She  would \emph{not}  turn  her  detector at all if the  answer is ``no". Bob  will then  monitor the  value of the expansion of  nearby   geodesics (which might be hard technically  as  he is not moving along one), by  say looking at freely falling particles  he is continuing releasing.  Bob  would have to be   very careful  to ensure that nothing he  does  generates in his  surroundings  any   energy momentum tensor that could  mimic  that  associated   with the change in the  state resulting from  a  detection  of a quantum field by  Alice's  detector  (or to take any energy momentum he generates  into account so as  to be able to distinguish that from  the one associated with the latter). If at any time in his  world line  he  detects   such   geodesic  expansion  he would know Alice's  decision is ``yes". The point  is that no mater when along  his world-line that would  happen the communication would be  superluminal as Alice and Bob are never in causal contact. This is of course  a rather poor signaling protocol because  Bob could eventually know if  Alice's decision is ``yes"  but  he would never know if Alice's decision is ``no".

 We can however remedy this by having Alice, say, use two different kind of detectors (with say vastly different energy of excitation, say one of then being tenfold, or having  different orientation), which could produce two different changes in the energy momentum tensor at  Bob's location. The  detection of  one  of the two  changes   would mean the  answer is  ``yes"  and the other  the  answer is ``no".  This  would  correspond to a complete 1- bit signaling. 
   
  Again the  efficiency  and reliability of this  protocol depends strongly on the  magnitude of  Alice's  and Bob absolute accelerations  $a_{\rm A}$ and  $ a_{\rm B}$  as well  as  the change on the   state of the quantum field induced by the  excitation  of   Alice's  detector (which the timing  we  recall  she cannot   control,  as  she can only decide  whether to turn or not the detector on). This change   would depend also  on Alice  detector's   which we can modify  by  selecting,  say, the energy gap, $\Omega$, with which the detector  operates.  
   
Ignoring switching effects, reasonably good performance for the detector could be achieved by setting the energy gap near the Unruh temperature (in natural units), since for long-time interactions the transition probability of the detector should behave as a Planckian distribution \cite{Fewster:2016ewy}. This  part can be modified  and we see  no obstacle,  in principle,  to make  this  quantity  arbitrarily large.  At the  same time  the   distance  $D_{\rm A}$  along  a hypersurface  orthogonal to $\xi$  from Alice's intersection  world-line and  say the bifurcating  surface of the  Killing Horizon is   determined  by  $a_{\rm A}$   but   decreases   as  the later increases   it, so for larger $D_{\rm A}$ the protocol becomes poorer.
   
   Optimizing    the protocol performance   with respect to $ a_ {\rm B}$ presumably  will   have to balance the  fact that   the analogous  distance  from  Bob  to the bifurcating horizon decreases with increasing  $a_{\rm B}$ and the detailed functioning of the device  he   uses in  detecting the expansion of nearby  geodesics  (an  analysis that   does not seem   very easy to carry out).

   In  any  event  we think  that  even if the optimal level of  efficiency is not  very high,  that does not detract from the fact that, in principle,    there  are  no trivial   or   evident   obstacles  for the protocol to  work at  some level of reliability  and any nonzero value of that  would  represent  the opening of a door for superluminal  communication. Indeed, there exist currently serious proposals to measure the gravitational field with high precision, e.g. \cite{SmallGrav}.  We will however  discuss  further down the   manuscript what  seem to be  the most  natural possibilities for nature to  prevent,   in principle,   the  working  of such protocol. 

In  any   event it seems that, even if not  highly effective, the possibility of   Alice  signaling to Bob   in a  superluminal way   should  be considered  problematic in view  of the implications  that would have  for    our understanding of the world. In fact, the situation is aggravated by the fact that even if the detector does not detect a particle, there are higher-order effects (in the coupling constant) that induce changes in the stress-energy tensor on the left Rindler wedge \cite{AudMul}.\footnote{Although it is not clear if the effect when no particle is absorbed is an artifact of perturbation theory.}

Thus, it seems imperative to consider the caveats that might provide a path to avoid such a problematic communication protocol. Before we   start let us   make some  observations   regarding   semiclassical gravity.    Semiclassical gravity is   a formalism in which the gravitational field is taken as  dynamical but  treated in classical terms as the metric of a spacetime $(M,  g_{ab})$  in general relativity, while the  matter is treated in the language of quantum fields on such spacetime. The metric is taken to  satisfy  the  semiclassical  version of  Einstein's field equations
\begin{equation}
\label{semi-simple}
G_{ab} (x) = 8\pi G  \langle T^{\rm ren}_{ab} (x) \rangle,
\end{equation} 
with $\langle T_{ab}^{\rm ren} \rangle $, the expectation value of the matter's quantum stress-energy tensor in a suitable quantum state, while the quantum matter obeys the dynamics of QFT. See e.g. \cite{QGFord} for a review.

In incorporating state collapses in the semiclassical gravity context we remark that there is an issue, which we have touched on briefly above. The point is  that once  something  like  state reduction is considered  (be it in the Copenhaguen approach or  in any  of the spontaneous  collapse theories models, e.g. \cite{collapse theories, collapse theories-2, collapse theories-3, collapse theories-4, collapse theories-5, collapse theories-6, collapse theories-7, collapse theories-8, collapse theories-9, collapse theories-10, collapse theories-11, collapse theories-12}),   one has the ambiguity of which  state  should be used in computing the  right-hand side  of eq. \eqref{semi-simple} should be employed.  In other words, it seems  natural that if we  are interested in the value of the left-hand side of eq. \eqref{semi-simple} at a point  $x$, the right-hand side  should  be computed using a state associated  with a Cauchy  surface that passes  trough   $x$ ,  but of  course   there  are infinite  such surfaces and  the  proposal to consider  eq. \ref{semi-simple}, even as an approximation,  must be completed  with a detailed  prescription  in this regard   in order to at  least have a  well-defined proposal \cite{Hypersurface, Hypersurface-2, Hypersurface-3, Hypersurface-4, Hypersurface-5}.  We should emphasize however that no matter what Cauchy surface is used, a portion of the Cauchy surface must extend to the left Rindler wedge. Thus, assuming that the collapse occurs along any arbitrary Cauchy surface, the left Rindler wedge must include a ``pre-collapse" and a ``post-collapse" region with different spacetime geometry -- according to semiclassical gravity.

Let us now offer,  and  briefly discuss what we think is the list of  serious  set of options to be considered, that could help in avoiding  the conclusion of super-luminal signaling:

\begin{itemize}
\item[(i)] {\it  Strong  enough departures from semiclassical gravity, at least  in region II}:
\item[(ii)] {\it Existence of effects that are indistinguishable from observations of the change in  the stress-energy  in region II}:
\item[(iii)] {\it A fundamental indetectability of the change of the state in  region II,  and in particular  that of  expectation value of the stress-energy tensor in said  region}:
\item[(iv)]  {\it   Some  fundamental  impediment  of the  construction of the    set up.}
\end{itemize}

 Let us now   briefly   discuss  the options  (i)-(iv) considered above. 
  
 (i)  First, and   in order to  address the   issue,  we must  clarify  what is meant by the words ``strong enough".  We take that to  indicate that  as  a  result of  some unknown aspect of  physics (with originating, say, in aspects  of  quantum gravity), the validity of eq. \eqref{semi-simple} with the right-hand side in the updated state, cf. the discussion of sec. \ref{sec:ChangeTab}, would  be  violated to such a large degree that,   for any    design  of Bob  measuring  instrument, the expected  result  will be  modified  by a factor  of at least the   same  order of magnitude. As we have pointed out in sec. \ref{sec:ChangeTab}, it is possible to make the expectation value of the stress-energy tensor at points in  region II \footnote{It is worth noting that our faster-than-light signaling protocol does not really require the validity of semiclassical gravity during the measurements. The only real requirement is that Bob be able to measure every once in a while the expectation value of the energy momentum tensor by any means. As such, this result seems to contradict the so called no signaling theorems. However, the particular experimental situation we are considering does not fall under the hypothesis of such theorems which concerns finite time measurements. We will discuss this issue further in this paper.} arbitrarily large in the post-collapse state by increasing the detector gap. However, it must be acknowledged that if $\Omega$ becomes large enough modeling the state collapse in semiclassical gravity along a Cauchy surface could be questioned on the basis that a large violation of the stress-energy tensor along a Cauchy surface is too strong a deviation from the semiclassical regime. Thus,  one might  argue  that although one could trust semiclassical gravity before and after the state collapse, there is no way to model the system ``during" the state collapse in semiclassical terms, and a more refined understanding of how to model state collapses in semiclassical gravity could be required. We note  however  that even   if that is   the case,  such   conclusion  will not prevent the   super-luminal signaling as  all we  need  is for Bob to detect   the   effect at any time. 
 
  On the other hand, it  seems   rather unlikely that arbitrarily large  departures  of   semiclassical gravity  will be  associated  with    such simple and  rather common situation.  The  full  analysis  of  the question  thus requires consideration of the means  by which  Bob might  detect the  corresponding  change in  the  spacetime metric in  region II \footnote{ We will not study here in detail the detection method to be  employed by Bob,    but   just point out  that interesting  ideas  to  measure  small gravitational  effects  have been used  in say   searches for  deviation of the universality  of free fall, and  also  note a recent proposl   to look  for  Plank scale   dark matter \cite{DarkMaterSerach}. }.   As it seems  clear that a   modification of the spacetime curvature  might,   in principle, be   measured   by the study of the geodesic deviation equation   and  in  particular the expansion of the congruence of  geodesics  in that  region, it appears  that  the   question will have to be connected at  least  to  some fundamental  limitation on the  validity of the notion of test   point particles   following  geodesics of the underlying metric.  There are of  course  some  limitations  on that arising from simple quantum  considerations  about  the  description of so called  free  point particles  which in fact negate  the notion of well defined trajectories.   
  
     Moreover,  as  discussed in \cite{Chryss}, the  standard quantum mechanical  minimal   de-localization of   a quantum particle  (characterized  for instance  by  its  Compton  wavelength)  implies  such particle  ought not  to  be  considered  as  point like,  and as is well known,   in general,  even at the classical level,  extended  objects  fail to follow  geodesics.  It is  unclear  at this point if these  kind of considerations will  be  enough to dismiss  our example, given the fact that, as noted, the effect could be  made as  large as  one  wants and the time  available  for  Bob to make the measurement is  arbitrarily large. 
     
   Finally, one can raise the issue of whether one should not  trust  semiclassical gravity in any situation  in which the quantum fluctuations (i.e. the  quantum uncertainties ) in the   energy momentum tensor are larger  in magnitude than the  expectation value of  the energy momentum tensor.  Considering that such a restriction  would imply that semiclassical gravity  can not be used in  the case of Minkowski vacuum, it seems to us that  the use of such a generic prohibition  would rule out the  use of semi-classical gravity in essentially all situations. That, it seems to us, would be  a very  drastic and unwarranted conclusion. 
    
  (ii)  Here   we   must consider   the   existence of other  effects  that might not be   effectively distinguished   from the  changes  resulting from the collapse. Those  might be  intrinsically associated  with the  gravitational effects of   either  the measuring  devices  that Bob  would  be introducing in order to detect the  changes in the   spacetime  metric  in  region II. They also might be associated  with mere  existence of  (the non-vanishing   stress-energy tensor corresponding to)  Alice and her detectors,  whose own gravitational effects  we have  been neglecting so  far, although classically, these effects will in principle propagate causally and not affect region II. It is likely that deviations from Minkowski spacetime reflect on the \emph{non-local} state of the field from a semiclassical-gravity viewpoint in such a way that the quantum state of the field cannot be in principle the Minkowski vacuum state to begin with. Moreover, the stress-energy of Alice and her apparatus must be considered in the constraints of the theory, and the spacetime will have a non-vanishing ADM mass, reflected in the asymptotic behavior in all spacetime directions.
  
Another possibility is to consider that the model used for Alice's detector, although standard in the literature, is a local model (in Region I) of the detector. Treating Alice's detector as a quantum field yields a detector model with support in both Regions I and II. In the Heisenberg picture we know that the existence of a field-like detector for Alice will only affect the dynamics in the causal future of the coupling region between Alice's detector (or probe) and the quantum field considered as a system. This can also be formalised in the algebraic QFT language \cite{Fewster:2018qbm}. However, as emphasised in \cite{Fewster:2018qbm}, a local and covariant measurement scheme is only able to describe the \emph{probe-system measurement chain} under the assumption that somewhere, someone knows how to measure something. Thus, it seems that the issue of sending a faster-than-light signal by the ``act of measurement" cannot be resolved by simply changing the detector model. Sorkin has pointed out in his ``impossible measurements" protocol \cite{Sorkin Imposible} that the  existence of spatially extended  detectors,  would  make  faster than light signaling  almost unavoidable. See however \cite{Bostelmann:2020unl}.

  (iii)   This  possibility    could  result    from  a  variety of reasons.   For instance,  it might be  that  the  appropriate  versions of semiclassical gravity   that   will hold are such that the   expectation value of the stress-energy tensor at any point,  is taken on   states associated with hypersurfaces  which  would   not incorporate the change in the state  resulting from a collapse  or a measurement on region  I.  
  One  such  option   would  require  to  take the  expectation value  of the stress-energy tensor  appearing   at the right-hand side of  Einstein's  semiclassical equation in  $x$ to be  computed  
  using  the   state corresponding to the something like  the  hypersurface  $\partial J^{-} (x)$ (the  boundary of the causal past of $ x$), and  that, at the same time, the  detailed  theory characterising  the measurement  (i.e., something like  the   spontaneous  collapse theories  considered in  \cite{Collapse5, Collapse6, Collapse7}),  is such  that  the state associated with  $\partial J^{-} (x)$ is  unaffected.    A  scheme  of that  kind   would   ensure   that the measurement of Alice would have no consequences from semiclassical gravity for spacelike-separated events. Such proposal  is not without  difficulties, one  of which is the fact that in general $\partial J^{-} (x)$  is not a  Cauchy   surface,  and  another is  that is  not smooth   at $x$. The first difficulty might be resolved if one has  some good  reason to   consider that there is a certain  initial state associated with an initial   Cauchy  hypersurface  $ \Sigma_{in}$, which   might be  considered  as  the initially  ``{\it prepared} "   state,  or  perhaps  {\it  the initial state of the  universe}  or something like that, and then to   consider,  computing the expectation values of quantities of interest   at $x$,  using the  state associated  with the    surface $ ( I^{+} (\Sigma_{in}) \cap \partial J^{-} (x) )\cup (\Sigma_{in} - J^{-} (x)) $,   while the second  problem  might be dealt with an adjustment of  the recipe  based on the  taking of   appropriate limits of  a  succession of   smooth  Cauchy  hypersurfaces that   have  as a limit (in a suitable sense),  the hypersurface  indicated above. The issue  is  however  quite delicate as  illustrated by the   discussions in  \cite{Hypersurface}, the pursuit of which lies  outside the  scope of the present work.

An argument in favour of the impossibility of detecting the signal is the following. One can assume that stress-energy conservation must hold \emph{on average}, i.e., for successive measurements, which prevents one from extracting arbitrarily large amounts of energy from the quantum field by detector measurements, as argued in \cite{Unruh-Wald}. Thus, an arbitrarily-efficient detector must produce arbitrarily-small changes for the stress-energy tensor when it detects a Rindler particle. The issue now becomes whether there exists an ``engineering window" where the efficiency of the detector and the amplitude of the signal can be compensated, such that a faster-than-light signal is measurable by Bob and that such a signal can be sent with sufficient certainty by Alice. On this point, it follows from the inequality $\Delta E  \Delta T \geq  \hbar $ that to detect such a small signal Bob requires a large amount of time. On the other hand, in principle, Bob has an infinite time to measure an arbitrarily small signal, for he can orbit along a Killing vector boost orbit in the left Rindler wedge, and one could argue that such engineering window can always be found. However, any physical signal sent by Alice must decay as it approaches $\mathscr{I}^+$.Thus, it is  expected that the signal  will become weaker as Bob's proper time elapses,  and this might  render the signal  ``effectively undetectable ". 

 (iv) Finally there could be aspects of the set up that  would  make it   simply unfeasible.     One  possibility  we should   consider is the following.  In  order for  Bob to be sure that  the  change in the energy momentum he observes  corresponds indeed to a signal  that was  sent  by  Alice,  he must be sure that similar  signals  are  nor reaching him  coming from   elsewhere. That is,  he  must be  quite sure that the  state of the quantum field prior to Alice's manipulation of her  detector is  the Minkowski  vacuum, and  as he must be ready to receive Alice's   signal   at  any point in his  world line,  he  must be sure that  such  characterization of the state of the  quantum field  must be   the appropriate one up   to  arbitrarily  distant  regions of  ``space". So there must be a preparation of the initial state of the     quantum field  on an extremely large  spatial region occurring well before the whole protocol is  even  started (and if we  want  to ensure Bob  has in    effect  arbitrarily long time to make the detection   it seems the preparation of the   state ought to   encompass  a full  Cauchy hypersurface.  However, as  already noted, the   simultaneous (in some  reference frame)  measurements of  observables associated with such extended spatial regions -- say, as a means of state preparation -- are known to be rather problematic in their own right as  argued by Sorkin\cite{Sorkin Imposible} (again, see however \cite{Bostelmann:2020unl}). 
 
  A related    difficulty might reside in the very preparation of the setup which in  our case  starts  with ensuring the initial state of the quantum  filed  is the Minkowski  vacuum. As   an important feature of our example is the fact that   the   superluminal signaling  protocol  is  expected to  work    considering arbitrarily large  values of the  proper times  at which Alice's detector clicks  and    at  which Bob  makes his observation, it is imperative that  the Minkowski vacuum    characterized the  state of the field not just as a local approximation but in an  arbitrarily large  region,  otherwise  Bob  might not be  able to tell if what he detected is a result of  Alice's choice of turning her  detector on, or some other perturbation  generated  elsewhere.   Such a global preparation  seems to contain  non-local elements that are similar to those  occurring in   Sorkin's  Impossible measurements  example, and might be  subjected  to  similar  questionings. Furthermore, one might argue  that  as we need to  include Alice , Bob and  their  measuring devices then, the state of the field  could not possibly be, strictly  speaking, the Minkowski vacuum.  It would be  surprising  if such concerns could not be   overcome  even in principle.   Needles is to  say that a  rather general  argument along such lines  is not available at this time as far as we know. 

In any case,   such   arguments   must be considered  with care,   because in principle,   even a  very  unreliable   communication protocol which allows faster than light  communications  would be  quite problematic.  We  think these ideas deserve much deeper  exploration.

\section{Final remarks}      \label{sec:Conclusions}

 We have revisited  the issue  of  detection of  a Rindler particle  by  an accelerator detector confined to the right Rindler wedge, focusing   attention on the  implications of the reduction of the  state  associated  with the actual detection or  what is often  referred to  as the measurement part of the process. We have  noted that the concomitant  modification of the state of the quantum  field, imply changes in the (expectation value of the renormalised) stress-energy tensor in both Rindler wedges.
 
Concerning the right Rindler wedge, it is interesting that the  resulting state  and  stress-energy tensor expectation values    become non-thermal  which  is,  in a  sense,  easy to  understand as  result of the disruption brought by the interaction with the detector. Of course, one  expects   that  in the long run further interaction  between the field and  detector or  among the field modes themselves will bring the  system to a new   state of thermal equilibrium.

Concerning the left Rindler wedge, however, the change in the stress-energy tensor expectation value is more problematic, as it seems to open a possibility for  faster  than light communication.  We have noted  some  of what  we see  as the most natural   possibilities to  avoid such  conclusion, but further studies of these issues  are required  in  order to   get  to  more definite  and solid conclusions.

For the moment, we can only stress that the fuziness in the  theoretical characterization of the  act of measuring in quantum theory can bring about complications even in apparently inocuous circumstances, such as in the context of the detection of a Rindler particle in Minkowski spacetime. Despite being overlooked in many practical applications in physics, the literature on the measurement problem is quite large, and the positions taken in its  face are quite varied. We do not intend to discuss them in detail. It suffices for us to mention that the paths to overcome the measurement problem have been classified by Maudlin \cite{TimTopoi} as follows: It is internally inconsistent to hold simultaneously the following three propositions about a quantum theory:
\begin{enumerate}[(i)]
\item The description of a physical system as provided by the quantum state or wave function is complete.
\item The evolution of the  quantum state is always dictated by the Sch\"rodinger equation (or its relativistic generalisations).
\item  Individual experiments produce definite (even if often unpredictable) results.
\end{enumerate}

Thus, one must negate at least one of (i)-(iii) above. Negating (i) leads down, in general, the path of the so-called {\it hidden variable theories}, of which the de Broglie-Bohm theory is the best known example \cite{Bohm}.  Negating (ii) implies that the  collapse or reduction of the wave function plays a central r\^ole, as in the Copenhaguen textbook interpretation, as well as in so-called {\it spontaneous  collapse} or {\it dynamical state reduction} theories, such as GRW, CSL, Penrose-Di\'osi, etc \cite{Collapse1,Collapse2, Collapse3, Collapse4, Collapse5, Collapse6, Collapse7}. The negation of (iii) leads to {\it many-worlds}- or {\it many-minds}-type interpretations \cite{ManyWorlds1, ManyWorlds2}. 

 In any case, we should restate that the superluminar tension relies on the application of two central hypotheses in this work. The first one is that measurements induce the collapse of the wavefunction, and that such collapse occurs on a Cauchy surface of spacetime. The second one is that the problem studied in (nearly) flat spacetime is within the regime of applicability of semiclassical gravity. It is well possible that either one of the two hypotheses fail, or that they cannot be taken to hold together. If this is the case, it would seem that there exist apparently inocuous situations, such as the one here studied, for which a correct theoretical description of the evolution of the system must resort to quantum gravity. This seems to be in agreement with conclusions drawn from \cite{Belenchia:2019gcc, Wald:2020jow}. However if that is the case it seems  likely  that other situations that  are often   discussed  using semiclassical  language,  for instance  Hawking  radiation and the backreaction in black holes (even  at early times),   might require reconsideration as  well.  

Finally, it is our hope to draw the attention of the community to these delicate issues that afflict our understanding of quantum theory in general and of QFT in particular, especially in gravitational contexts.

\section*{Acknowledgements} 
We  gladly  thank  Prof. George Matsas for  his  careful reading of a previous version of this  work and     very useful comments, and Prof. Stephen A. Fulling  for very stimulating discussions, as well as  Ruth E. Kastner for pointing out a few mistakes in the  original version of this  work.  B.A.J-A. is supported by a CONACYT postdoctoral fellowship. D.S. acknowledges partial financial support from  PAPIIT-UNAM, Mexico (Grant No.IG100120); the Foundational Questions Institute (Grant No. FQXi-MGB-1928); the Fetzer Franklin Fund, a donor advised by the Silicon Valley Community Foundation. B.A.J.-A. and D.S. acknowledge the support of CONACYT grant  FORDECYT-PRONACES no. 140630. 

\appendix 

\section{The Minkowski vacuum in the Rindler wedges}
\label{App:MinkoVac} 
 
The initial state is taken to be the Minkowski vacuum, whose Wightman function takes the following form in the right Rindler wedge
\begin{align}
\langle \Omega_{\rm M} | \hat \Phi(x) \hat \Phi(x') \Omega_{\rm M} \rangle = \int_{\mathbb{R}^+\times \mathbb{R}^2}  \!\!\!\!\!\!\!\!\!\! \dd^3\kappa \int_{\mathbb{R}^+ \times \mathbb{R}^2} \!\!\!\!\!\!\!\!\!\! \dd^3 \kappa' \langle \Omega_{\rm M} | \left( v_{I \vec \kappa}(x) \hat a_{{\rm R} \vec \kappa} + \overline{v_{I \vec \kappa}(x)} \hat a^*_{{\rm R} \vec \kappa}  \right) \left( v_{I \vec \kappa'	}(x') \hat a_{{\rm R} \vec \kappa'} + \overline{v_{I \vec \kappa'}(x')} \hat a^*_{{\rm R} \vec \kappa'}  \right) \Omega_{\rm M} \rangle.
\end{align}

Using the relations \cite[eq. 2.125-2.127]{Crispino},
 \begin{subequations}
 \begin{align}
 \langle \Omega_{\rm M} |  \hat a^*_{{\rm R} \vec \kappa}  \hat a_{{\rm R} \vec \kappa'}  \Omega_{\rm M} \rangle  = \langle \Omega_{\rm M} |  \hat a^*_{{\rm L} \vec \kappa}  \hat a_{{\rm L} \vec \kappa'}  \Omega_{\rm M} \rangle & = \left( e^{2 \pi \omega/a} -1\right)^{-1} \delta^3(\vec \kappa - \vec \kappa'), \\
 \langle \Omega_{\rm M} | \hat a_{{\rm R} \vec \kappa}  \hat a^*_{{\rm R} \vec \kappa'}   \Omega_{\rm M} \rangle  = \langle \Omega_{\rm M} | \hat a_{{\rm L} \vec \kappa}  \hat a^*_{{\rm L} \vec \kappa'}   \Omega_{\rm M} \rangle & = \left(1- e^{-2 \pi \omega/a}\right)^{-1} \delta^3(\vec \kappa - \vec \kappa') , \\
  \langle \Omega_{\rm M} |  \hat a_{{\rm L} \vec \kappa}  \hat a_{{\rm R} \vec \kappa'}  \Omega_{\rm M} \rangle  = \langle \Omega_{\rm M} |  \hat a^*_{{\rm L} \vec \kappa}  \hat a^*_{{\rm R} \vec \kappa'}  \Omega_{\rm M} \rangle & = \left( e^{\pi \omega/a} - e^{-\pi \omega/a}\right)^{-1} \delta^3(\vec \kappa - \vec \kappa'),
 \end{align}
 (and zero otherwise) we obtain that
 \end{subequations}
\begin{align}
\langle \Omega_{\rm M} | \hat \Phi(x) \hat \Phi(x') \Omega_{\rm M} \rangle & =  \int_{\mathbb{R}^+ \times \mathbb{R}^2} \!\!\!\!\!\!\!\!\!\! \dd^3 \kappa  \left( \frac{v_{I \vec \kappa}(x)\overline{v_{I \vec \kappa}(x')}}{1- e^{-2 \pi \omega/a}} + \frac{\overline{v_{I \vec \kappa}(x)}  v_{I \vec \kappa	}(x')}{e^{2 \pi \omega/a}-1} \right).
\label{MinkoR}
\end{align}

Likewise, in the left Rindler wedge the Wightman function takes the form
\begin{align}
\langle \Omega_{\rm M} | \hat \Phi(x) \hat \Phi(x') \Omega_{\rm M} \rangle & =  \int_{\mathbb{R}^+ \times \mathbb{R}^2} \!\!\!\!\!\!\!\!\!\! \dd^3 \kappa  \left( \frac{v_{II \vec \kappa}(x)\overline{v_{II \vec \kappa}(x')}}{1- e^{-2 \pi \omega/a}} + \frac{\overline{v_{II \vec \kappa}(x)}  v_{II \vec \kappa	}(x')}{e^{2 \pi \omega/a}-1} \right),
\label{MinkoL}
\end{align}
where $v_{II \vec \kappa}$ are left Rindler modes.
 
\section{Normalisation of the updated state} 
\label{App:Normalisation}
 
We have seen in sec. \ref{sec:Main} that if the detector clicks the updated state of the field becomes
\begin{align}
|f\rangle = -i \mathcal{N} \int_{\mathbb{R}^+ \times \mathbb{R}^2} \!\!\!\!\!\!\!\!\!\! \dd^3 \kappa \int_{I} \dd \vol(x) \zeta(x) \left( v_{I \vec \kappa}(x) \hat a_{{\rm R} \vec \kappa} + \overline{v_{I \vec \kappa}(x)} \hat a^*_{{\rm R} \vec \kappa}  \right) |\Omega_{\rm M}\rangle.
\end{align} 

In this appendix, we see that the normalisation constant is given by eq. \eqref{Normalisation}. 
We compute
\begin{align}
\mathcal{N}^{-2} \langle f|f \rangle & =  \int_{\mathbb{R}^+ \times \mathbb{R}^2} \!\!\!\!\!\!\!\!\!\! \dd^3 \kappa \int_{\mathbb{R}^+ \times \mathbb{R}^2} \!\!\!\!\!\!\!\!\!\!  \dd^3 \kappa' \int_{I} \dd \vol(x)  \int_{I} \dd \vol(x') \overline{\zeta(x)} \zeta(x') \nonumber \\
& \times \langle \Omega_{\rm M} | \left( v_{I \vec \kappa}(x) \hat a_{{\rm R} \vec \kappa} + \overline{v_{I \vec \kappa}(x)} \hat a^*_{{\rm R} \vec \kappa}  \right) \left( v_{I \vec \kappa'	}(x') \hat a_{{\rm R} \vec \kappa'} + \overline{v_{I \vec \kappa'}(x')} \hat a^*_{{\rm R} \vec \kappa'}  \right) \Omega_{\rm M} \rangle.
\end{align} 

Using the relations \cite[eq. 2.125-2.126]{Crispino}, we obtain that

 \begin{align}
\mathcal{N}^{-2} \langle f|f \rangle & =  \int_{\mathbb{R}^+ \times \mathbb{R}^2} \!\!\!\!\!\!\!\!\!\! \dd^3 \kappa  \int_{I} \dd \vol(x)  \int_{I} \dd \vol(x') \overline{\zeta(x)} \zeta(x') \left( \frac{v_{I \vec \kappa}(x)\overline{v_{I \vec \kappa}(x')}}{1- e^{-2 \pi \omega/a}} + \frac{\overline{v_{I \vec \kappa}(x)}  v_{I \vec \kappa	}(x')}{e^{2 \pi \omega/a}-1} \right).
\label{N-2} 
 \end{align}
 
 \section{The two-point function in the updated state in the right Rindler wedge}
\label{Sec:RightUpdate}

The updated stress-energy tensor can be computed from the two-point function in the updated state. In this appendix, we show that the two-point function in the updated state in the right Rindler wedge is given by
\begin{align}
\langle f | \hat \Phi(x) \hat \Phi(x') f \rangle & = \langle \Omega_{\rm M} | \hat \Phi(x) \hat \Phi(x') \Omega_{\rm M} \rangle+ \Delta_{\rm R}(x,x') .
\end{align}
where $\Delta_{\rm R}(x,x')$ is given by eq. \eqref{DeltaSimp} below.
 
The two-point function in the right Rindler wedge reads
\begin{align}
\langle f | \hat \Phi(x) \hat \Phi(x') f \rangle & = \mathcal{N}^{2} \int_{\mathbb{R}^+ \times \mathbb{R}^2} \!\!\!\!\!\!\!\!\!\! \dd^3 \kappa \int_{\mathbb{R}^+ \times \mathbb{R}^2} \!\!\!\!\!\!\!\!\!\! \dd^3 \kappa'  \int_{\mathbb{R}^+ \times \mathbb{R}^2} \!\!\!\!\!\!\!\!\!\! \dd^3 p \int_{\mathbb{R}^+ \times \mathbb{R}^2} \!\!\!\!\!\!\!\!\!\!  \dd^3 p' \int_{I} \dd \vol(y)  \int_{I} \dd \vol(y') \overline{\zeta(y)} \zeta(y') \nonumber \\
& \times \langle \Omega_{\rm M} | \left( v_{I \vec p}(y) \hat a_{{\rm R} \vec p} + \overline{v_{I \vec p}(y)} \hat a^*_{{\rm R} \vec p}  \right) \left( v_{I \vec{\kappa}}(x) \hat{a}_{{\rm R} \vec{\kappa}} + \overline{v_{I \vec{\kappa}}}(x) \hat{a}_{{\rm R} \vec{\kappa}}^* \right) \nonumber \\
& \times \left( v_{I \vec{\kappa}'}(x') \hat{a}_{{\rm R} \vec{\kappa}'} + \overline{v_{I \vec{\kappa}'}}(x') \hat{a}_{{\rm R} \vec{\kappa}'}^* \right) \left( v_{I \vec p'	}(y') \hat a_{{\rm R} \vec p'} + \overline{v_{I \vec p'}(y')} \hat a^*_{{\rm R} \vec p'}  \right) \Omega_{\rm M} \rangle.
\label{2pt-f}
\end{align} 

Using \cite[eq. 2.122-2.124]{Crispino} one can obtain the relations
\begin{subequations}
\begin{align}
& \langle \Omega_{\rm M} | \hat{a}_{{\rm R} \vec{\kappa_1}} \hat{a}_{{\rm R} \vec{\kappa_2}} \hat{a}_{{\rm R} \vec{\kappa_3}} \hat{a}_{{\rm R} \vec{\kappa_4}} \Omega_{\rm M} \rangle = 0, \nonumber \\
& \langle \Omega_{\rm M} | \hat{a}_{{\rm R} \vec{\kappa_1}} \hat{a}_{{\rm R} \vec{\kappa_2}} \hat{a}_{{\rm R} \vec{\kappa_3}} \hat{a}_{{\rm R} \vec{\kappa_4}}^* \Omega_{\rm M} \rangle  = 0,\\
& \langle \Omega_{\rm M} | \hat{a}_{{\rm R} \vec{\kappa_1}} \hat{a}_{{\rm R} \vec{\kappa_2}} \hat{a}_{{\rm R} \vec{\kappa_3}}^* \hat{a}_{{\rm R} \vec{\kappa_4}} \Omega_{\rm M} \rangle  = 0,\\
&\langle \Omega_{\rm M} | \hat{a}_{{\rm R} \vec{\kappa_1}} \hat{a}_{{\rm R} \vec{\kappa_2}}^* \hat{a}_{{\rm R} \vec{\kappa_3}} \hat{a}_{{\rm R} \vec{\kappa_4}} \Omega_{\rm M} \rangle  = 0,\\
&\langle \Omega_{\rm M} | \hat{a}_{{\rm R} \vec{\kappa_1}}^* \hat{a}_{{\rm R} \vec{\kappa_2}} \hat{a}_{{\rm R} \vec{\kappa_3}} \hat{a}_{{\rm R} \vec{\kappa_4}} \Omega_{\rm M} \rangle  = 0,\\
&\langle \Omega_{\rm M} | \hat{a}_{{\rm R} \vec{\kappa_1}} \hat{a}_{{\rm R} \vec{\kappa_2}} \hat{a}_{{\rm R} \vec{\kappa_3}}^* \hat{a}_{{\rm R} \vec{\kappa_4}}^* \Omega_{\rm M} \rangle  \nonumber \\
& = \frac{1}{1- e^{-2 \pi \omega_1/a}} \frac{1}{1- e^{-2 \pi \omega_2/a}} \left( \delta(\vec \kappa_1 - \vec \kappa_3) \delta(\vec \kappa_2 - \vec \kappa_4) + \delta(\vec \kappa_1 - \vec \kappa_4) \delta(\vec \kappa_2 - \vec \kappa_3) \right),\\
&\langle \Omega_{\rm M} | \hat{a}_{{\rm R} \vec{\kappa_1}} \hat{a}_{{\rm R} \vec{\kappa_2}}^* \hat{a}_{{\rm R} \vec{\kappa_3}} \hat{a}_{{\rm R} \vec{\kappa_4}}^* \Omega_{\rm M} \rangle = \frac{1}{1- e^{-2 \pi \omega_1/a}} \frac{1}{1- e^{-2 \pi \omega_3/a}}  \delta(\vec \kappa_1 - \vec \kappa_2) \delta(\vec \kappa_3 - \vec \kappa_4)  \\
& + \frac{1}{1- e^{-2 \pi \omega_1/a}} \frac{ e^{-2 \pi \omega_3/a}}{1- e^{-2 \pi \omega_3/a}}  \delta(\vec \kappa_1 - \vec \kappa_4) \delta(\vec \kappa_2 - \vec \kappa_3), \\
&\langle \Omega_{\rm M} | \hat{a}_{{\rm R} \vec{\kappa_1}}^* \hat{a}_{{\rm R} \vec{\kappa_2}} \hat{a}_{{\rm R} \vec{\kappa_3}} \hat{a}_{{\rm R} \vec{\kappa_4}}^* \Omega_{\rm M} \rangle  \\
& = \frac{e^{-2 \pi \omega_1/a}}{1- e^{-2 \pi \omega_1/a}} \frac{1}{1- e^{-2 \pi \omega_4/a}} \left(  \delta(\vec \kappa_1 - \vec \kappa_2) \delta(\vec \kappa_3 - \vec \kappa_4) + \delta(\vec \kappa_1 - \vec \kappa_3) \delta(\vec \kappa_2 - \vec \kappa_4) \right), \\
&\langle \Omega_{\rm M} | \hat{a}_{{\rm R} \vec{\kappa_1}} \hat{a}_{{\rm R} \vec{\kappa_2}}^* \hat{a}_{{\rm R} \vec{\kappa_3}}^* \hat{a}_{{\rm R} \vec{\kappa_4}} \Omega_{\rm M} \rangle  \\
& = \frac{1}{1- e^{-2 \pi \omega_1/a}} \frac{\ee^{-2 \pi \omega_4/a}}{1- e^{-2 \pi \omega_4/a}}  \left( \delta(\vec \kappa_1 - \vec \kappa_2)  \delta(\vec \kappa_3 - \vec \kappa_4) +  \delta(\vec \kappa_1 - \vec \kappa_3) \delta(\vec \kappa_2 - \vec \kappa_4)\right), \\
&\langle \Omega_{\rm M} | \hat{a}_{{\rm R} \vec{\kappa_1}}^* \hat{a}_{{\rm R} \vec{\kappa_2}} \hat{a}_{{\rm R} \vec{\kappa_3}}^* \hat{a}_{{\rm R} \vec{\kappa_4}} \Omega_{\rm M} \rangle = \frac{e^{-2 \pi \omega_1/a}}{1- e^{-2 \pi \omega_1/a}} \frac{\ee^{-2 \pi \omega_3/a}}{1- e^{-2 \pi \omega_3/a}}  \delta(\vec \kappa_1 - \vec \kappa_2) \delta(\vec \kappa_3 - \vec \kappa_4) \\
&  + \frac{\ee^{-2 \pi \omega_1/a}}{1- e^{-2 \pi \omega_1/a}} \frac{1}{1- e^{-2 \pi \omega_2/a}}   \delta(\vec \kappa_1 - \vec \kappa_4) \delta(\vec \kappa_2 - \vec \kappa_3), \\
&\langle \Omega_{\rm M} | \hat{a}_{{\rm R} \vec{\kappa_1}}^* \hat{a}_{{\rm R} \vec{\kappa_2}}^* \hat{a}_{{\rm R} \vec{\kappa_3}} \hat{a}_{{\rm R} \vec{\kappa_4}} \Omega_{\rm M} \rangle  \nonumber\\
& = \frac{e^{-2 \pi \omega_1/a}}{1- e^{-2 \pi \omega_1/a}} \frac{\ee^{-2 \pi \omega_2/a}}{1- e^{-2 \pi \omega_2/a}} \left(  \delta(\vec \kappa_1 - \vec \kappa_3) \delta(\vec \kappa_2 - \vec \kappa_4) +  \delta(\vec \kappa_1 - \vec \kappa_4) \delta(\vec \kappa_2 - \vec \kappa_3) \right),\\
&\langle \Omega_{\rm M} | \hat{a}_{{\rm R} \vec{\kappa_1}} \hat{a}_{{\rm R} \vec{\kappa_2}}^* \hat{a}_{{\rm R} \vec{\kappa_3}}^* \hat{a}_{{\rm R} \vec{\kappa_4}}^* \Omega_{\rm M} \rangle  = 0,\\
&\langle \Omega_{\rm M} | \hat{a}_{{\rm R} \vec{\kappa_1}}^* \hat{a}_{{\rm R} \vec{\kappa_2}} \hat{a}_{{\rm R} \vec{\kappa_3}}^* \hat{a}_{{\rm R} \vec{\kappa_4}}^* \Omega_{\rm M} \rangle  = 0,\\
&\langle \Omega_{\rm M} | \hat{a}_{{\rm R} \vec{\kappa_1}}^* \hat{a}_{{\rm R} \vec{\kappa_2}}^* \hat{a}_{{\rm R} \vec{\kappa_3}} \hat{a}_{{\rm R} \vec{\kappa_4}}^* \Omega_{\rm M} \rangle  = 0,\\
&\langle \Omega_{\rm M} | \hat{a}_{{\rm R} \vec{\kappa_1}}^* \hat{a}_{{\rm R} \vec{\kappa_2}}^* \hat{a}_{{\rm R} \vec{\kappa_3}}^* \hat{a}_{{\rm R} \vec{\kappa_4}} \Omega_{\rm M} \rangle  = 0,\\
&\langle \Omega_{\rm M} | \hat{a}_{{\rm R} \vec{\kappa_1}}^* \hat{a}_{{\rm R} \vec{\kappa_2}}^* \hat{a}_{{\rm R} \vec{\kappa_3}}^* \hat{a}_{{\rm R} \vec{\kappa_4}}^* \Omega_{\rm M} \rangle  = 0.
\end{align}
\label{creannR}
\end{subequations} 
 
Inserting eq. \eqref{creannR} into \eqref{2pt-f} we obtain that the two point function can be obtained as a sum of six contributions, each coming from one of the non-trivial expressions in eq. \eqref{creannR}, i.e., it takes the form
\begin{align}
\langle f | \hat \Phi(x) \hat \Phi(x') f \rangle(x,x') = \sum_{n = 1}^6 G_{{\rm R} n}(x,x').
\end{align}
Subtracting eq. \eqref{MinkoR} from this expression we obtain that
\begin{align}
\Delta_{\rm R}(x,x') = \langle f | \hat \Phi(x) \hat \Phi(x') f \rangle - \langle \Omega_{\rm M} | \hat \Phi(x) \hat \Phi(x') \Omega_{\rm M} \rangle = \sum_{n = 1}^6 \Delta_{{\rm R}n}(x,x')
\end{align}
with
\begin{align}
\Delta_{{\rm R}1}(x,x') & := \mathcal{N}^{2} \int_{\mathbb{R}^+ \times \mathbb{R}^2} \!\!\!\!\!\!\!\!\!\! \dd^3 \kappa   \int_{\mathbb{R}^+ \times \mathbb{R}^2} \!\!\!\!\!\!\!\!\!\! \dd^3 p  \int_{I} \dd \vol(y)  \int_{I} \dd \vol(y') \overline{\zeta(y)} \zeta(y')  \frac{v_{I \vec p}(y) v_{I \vec{\kappa}}(x) \overline{v_{I \vec{p}}}(x') \overline{v_{I \vec \kappa}(y')}}{(1-e^{-2\pi \omega_p/a}) (1-e^{-2\pi \omega_\kappa/a})}, \\
\Delta_{{\rm R}2}(x,x') & := \mathcal{N}^{2} \int_{\mathbb{R}^+ \times \mathbb{R}^2} \!\!\!\!\!\!\!\!\!\! \dd^3 \kappa  \int_{\mathbb{R}^+ \times \mathbb{R}^2} \!\!\!\!\!\!\!\!\!\! \dd^3 p  \int_{I} \dd \vol(y)  \int_{I} \dd \vol(y') \overline{\zeta(y)} \zeta(y')  \frac{v_{I \vec p}(y) \overline{v_{I \vec{p}}}(x) v_{I \vec{\kappa}}(x') \overline{v_{I \vec \kappa}(y')}}{(1- e^{-2 \pi \omega_p/a})(1- e^{-2 \pi \omega_{\kappa}/a})}, \\
\Delta_{{\rm R}3}(x,x') & :=  \mathcal{N}^{2}  \int_{\mathbb{R}^+ \times \mathbb{R}^2} \!\!\!\!\!\!\!\!\!\! \dd^3 \kappa \int_{\mathbb{R}^+ \times \mathbb{R}^2} \!\!\!\!\!\!\!\!\!\!  \dd^3 p \int_{I} \dd \vol(y)  \int_{I} \dd \vol(y') \overline{\zeta(y)} \zeta(y') \frac{\overline{v_{I \vec \kappa}(y)} v_{I \vec{\kappa}}(x) v_{I \vec{p}}(x') \overline{v_{I \vec p}(y')}e^{-2 \pi \omega_\kappa}}{(1-e^{-2 \pi \omega_\kappa})(1-e^{-2 \pi \omega_{p}})}  \nonumber \\
& + \mathcal{N}^{2}  \int_{\mathbb{R}^+ \times \mathbb{R}^2} \!\!\!\!\!\!\!\!\!\! \dd^3 \kappa \int_{\mathbb{R}^+ \times \mathbb{R}^2} \!\!\!\!\!\!\!\!\!\!  \dd^3 p \int_{I} \dd \vol(y)  \int_{I} \dd \vol(y') \overline{\zeta(y)} \zeta(y') \frac{\overline{v_{I \vec \kappa}(y)} v_{I \vec{p}}(x) v_{I \vec{\kappa}}(x') \overline{v_{I \vec p}(y')}e^{-2 \pi \omega_\kappa}}{(1-e^{-2 \pi \omega_\kappa})(1-e^{-2 \pi \omega_{p}})}, \\
\Delta_{{\rm R}4}(x,x') & := \mathcal{N}^{2}  \int_{\mathbb{R}^+ \times \mathbb{R}^2} \!\!\!\!\!\!\!\!\!\! \dd^3 \kappa \int_{\mathbb{R}^+ \times \mathbb{R}^2} \!\!\!\!\!\!\!\!\!\!  \dd^3 p \int_{I} \dd \vol(y)  \int_{I} \dd \vol(y') \overline{\zeta(y)} \zeta(y') \frac{v_{I \vec \kappa}(y) \overline{v_{I \vec{\kappa}}(x)} \overline{v_{I \vec{p}}(x')} v_{I \vec p}(y')e^{-2 \pi \omega_{p}}}{(1-e^{-2 \pi \omega_\kappa})(1-e^{-2 \pi \omega_{p}})}  \nonumber \\
& + \mathcal{N}^{2}  \int_{\mathbb{R}^+ \times \mathbb{R}^2} \!\!\!\!\!\!\!\!\!\! \dd^3 \kappa \int_{\mathbb{R}^+ \times \mathbb{R}^2} \!\!\!\!\!\!\!\!\!\!  \dd^3 p \int_{I} \dd \vol(y)  \int_{I} \dd \vol(y') \overline{\zeta(y)} \zeta(y') \frac{v_{I \vec \kappa}(y) \overline{v_{I \vec{p}}(x)} \overline{v_{I \vec{\kappa}}(x')} v_{I \vec p}(y')e^{-2 \pi \omega_{p}}}{(1-e^{-2 \pi \omega_\kappa})(1-e^{-2 \pi \omega_{p}})}, \\
\Delta_{{\rm R}5}(x,x') & := \mathcal{N}^{2}  \int_{\mathbb{R}^+ \times \mathbb{R}^2} \!\!\!\!\!\!\!\!\!\! \dd^3 \kappa \int_{\mathbb{R}^+ \times \mathbb{R}^2} \!\!\!\!\!\!\!\!\!\!  \dd^3 p \int_{I} \dd \vol(y)  \int_{I} \dd \vol(y') \overline{\zeta(y)} \zeta(y') \frac{\overline{v_{I \vec \kappa}(y)} v_{I \vec{\kappa}}(x) \overline{v_{I \vec{p}}(x')} v_{I \vec p}(y')e^{-2 \pi \omega_{\kappa}} e^{-2 \pi \omega_{p}}}{(1-e^{-2 \pi \omega_\kappa})(1-e^{-2 \pi \omega_{p}})}, \\
\Delta_{{\rm R}6}(x,x') & := \mathcal{N}^{2}  \int_{\mathbb{R}^+ \times \mathbb{R}^2} \!\!\!\!\!\!\!\!\!\! \dd^3 \kappa \int_{\mathbb{R}^+ \times \mathbb{R}^2} \!\!\!\!\!\!\!\!\!\!  \dd^3 p \int_{I} \dd \vol(y)  \int_{I} \dd \vol(y') \overline{\zeta(y)} \zeta(y') \frac{\overline{v_{I \vec \kappa}(y)} \overline{v_{I \vec{p}}(x)} v_{I \vec{\kappa}}(x') v_{I \vec p}(y') e^{-2 \pi \omega_{\kappa}} e^{-2 \pi \omega_{p}}}{(1-e^{-2 \pi \omega_\kappa})(1-e^{-2 \pi \omega_{p}})}.
\end{align}

Collecting, we can write
\begin{align}
\Delta_{\rm R}(x,x') & = \Delta_{{\rm R}1}(x,x') + \Delta_{{\rm R}3}(x,x') +\Delta_{{\rm R}5}(x,x') + {\rm c.c.},
\label{DeltaSimp}
\end{align}
which yields eq. \eqref{DeltaR}.

\section{The two-point function in the updated state in the left Rindler wedge}
\label{Sec:LeftUpdate}

The updated stress-energy tensor can be computed from the two-point function in the updated state. In this appendix, we show that the two-point function in the updated state in the right Rindler wedge is given by
\begin{align}
\langle f | \hat \Phi(x) \hat \Phi(x') f \rangle & = \langle \Omega_{\rm M} | \hat \Phi(x) \hat \Phi(x') \Omega_{\rm M} \rangle+ \Delta_{\rm L}(x,x'),
\end{align}
as in eq. \eqref{DeltaL}.

The two-point function in the left Rindler wedge reads
\begin{align}
\langle f | \hat \Phi(x) \hat \Phi(x') f \rangle & = \mathcal{N}^{2} \int_{\mathbb{R}^+ \times \mathbb{R}^2} \!\!\!\!\!\!\!\!\!\! \dd^3 \kappa \int_{\mathbb{R}^+ \times \mathbb{R}^2} \!\!\!\!\!\!\!\!\!\! \dd^3 \kappa'  \int_{\mathbb{R}^+ \times \mathbb{R}^2} \!\!\!\!\!\!\!\!\!\! \dd^3 p \int_{\mathbb{R}^+ \times \mathbb{R}^2} \!\!\!\!\!\!\!\!\!\!  \dd^3 p' \int_{I} \dd \vol(y)  \int_{I} \dd \vol(y') \overline{\zeta(y)} \zeta(y') \nonumber \\
& \times \langle \Omega_{\rm M} | \left( v_{I \vec p}(y) \hat a_{{\rm R} \vec p} + \overline{v_{I \vec p}(y)} \hat a^*_{{\rm R} \vec p}  \right) \left( v_{II \vec{\kappa}}(x) \hat{a}_{{\rm L} \vec{\kappa}} + \overline{v_{II \vec{\kappa}}}(x) \hat{a}_{{\rm L} \vec{\kappa}}^* \right) \nonumber \\
& \times \left( v_{II \vec{\kappa}'}(x') \hat{a}_{{\rm L} \vec{\kappa}'} + \overline{v_{II \vec{\kappa}'}}(x') \hat{a}_{{\rm L} \vec{\kappa}'}^* \right) \left( v_{I \vec p'	}(y') \hat a_{{\rm R} \vec p'} + \overline{v_{I \vec p'}(y')} \hat a^*_{{\rm R} \vec p'}  \right) \Omega_{\rm M} \rangle.
\label{2pt-fL}
\end{align} 

Using \cite[eq. 2.122-2.124]{Crispino} one can obtain the relations
\begin{subequations}
\begin{align}
& \langle \Omega_{\rm M} | \hat{a}_{{\rm R} \vec{\kappa_1}} \hat{a}_{{\rm L} \vec{\kappa_2}} \hat{a}_{{\rm L} \vec{\kappa_3}} \hat{a}_{{\rm R} \vec{\kappa_4}} \Omega_{\rm M} \rangle =  \frac{e^{- \pi \omega_1/a}}{1- e^{-2 \pi \omega_1/a}} \frac{e^{- \pi \omega_2/a}}{1- e^{-2 \pi \omega_2/a}}\tilde \delta(\vec \kappa_1 - \vec \kappa_3) \tilde \delta(\vec \kappa_2 - \vec \kappa_4) \nonumber \\
& + \frac{e^{- \pi \omega_1/a}}{1- e^{-2 \pi \omega_1/a}} \frac{e^{- \pi \omega_3/a}}{1- e^{-2 \pi \omega_3/a}}\tilde \delta(\vec \kappa_1 - \vec \kappa_2) \tilde \delta(\vec \kappa_3 - \vec \kappa_4), \nonumber \\
& \langle \Omega_{\rm M} | \hat{a}_{{\rm R} \vec{\kappa_1}} \hat{a}_{{\rm L} \vec{\kappa_2}} \hat{a}_{{\rm L} \vec{\kappa_3}} \hat{a}_{{\rm R} \vec{\kappa_4}}^* \Omega_{\rm M} \rangle  = 0,\\
& \langle \Omega_{\rm M} | \hat{a}_{{\rm R} \vec{\kappa_1}} \hat{a}_{{\rm L} \vec{\kappa_2}} \hat{a}_{{\rm L} \vec{\kappa_3}}^* \hat{a}_{{\rm R} \vec{\kappa_4}} \Omega_{\rm M} \rangle  = 0,\\
&\langle \Omega_{\rm M} | \hat{a}_{{\rm R} \vec{\kappa_1}} \hat{a}_{{\rm L} \vec{\kappa_2}}^* \hat{a}_{{\rm L} \vec{\kappa_3}} \hat{a}_{{\rm R} \vec{\kappa_4}} \Omega_{\rm M} \rangle  = 0,\\
&\langle \Omega_{\rm M} | \hat{a}_{{\rm R} \vec{\kappa_1}}^* \hat{a}_{{\rm L} \vec{\kappa_2}} \hat{a}_{{\rm L} \vec{\kappa_3}} \hat{a}_{{\rm R} \vec{\kappa_4}} \Omega_{\rm M} \rangle  = 0,\\
&\langle \Omega_{\rm M} | \hat{a}_{{\rm R} \vec{\kappa_1}} \hat{a}_{{\rm L} \vec{\kappa_2}} \hat{a}_{{\rm L} \vec{\kappa_3}}^* \hat{a}_{{\rm R} \vec{\kappa_4}}^* \Omega_{\rm M} \rangle  = \frac{1}{1- e^{-2 \pi \omega_1/a}} \frac{1}{1- e^{-2 \pi \omega_2/a}}\delta(\vec \kappa_1 - \vec \kappa_4) \delta(\vec \kappa_2 - \vec \kappa_3) \nonumber \\
& + \frac{e^{- \pi \omega_1/a}}{1- e^{-2 \pi \omega_1/a}} \frac{e^{- \pi \omega_3/a}}{1- e^{-2 \pi \omega_3/a}} \tilde \delta(\vec \kappa_1 - \vec \kappa_2) \tilde \delta(\vec \kappa_3 - \vec \kappa_4),\\
&\langle \Omega_{\rm M} | \hat{a}_{{\rm R} \vec{\kappa_1}} \hat{a}_{{\rm L} \vec{\kappa_2}}^* \hat{a}_{{\rm L} \vec{\kappa_3}} \hat{a}_{{\rm R} \vec{\kappa_4}}^* \Omega_{\rm M} \rangle = \frac{e^{- \pi \omega_1/a}}{1- e^{-2 \pi \omega_1/a}} \frac{e^{- \pi \omega_2/a}}{1- e^{-2 \pi \omega_2/a}}\tilde \delta(\vec \kappa_1 - \vec \kappa_3) \tilde \delta(\vec \kappa_2 - \vec \kappa_4) \nonumber \\
& + \frac{1}{1- e^{-2 \pi \omega_1/a}} \frac{e^{- 2 \pi \omega_2/a}}{1- e^{-2 \pi \omega_2/a}}\delta(\vec \kappa_1 - \vec \kappa_4) \delta(\vec \kappa_2 - \vec \kappa_3) \\
&\langle \Omega_{\rm M} | \hat{a}_{{\rm R} \vec{\kappa_1}}^* \hat{a}_{{\rm L} \vec{\kappa_2}} \hat{a}_{{\rm L} \vec{\kappa_3}} \hat{a}_{{\rm R} \vec{\kappa_4}}^* \Omega_{\rm M} \rangle = 0, \\
&\langle \Omega_{\rm M} | \hat{a}_{{\rm R} \vec{\kappa_1}} \hat{a}_{{\rm L} \vec{\kappa_2}}^* \hat{a}_{{\rm L} \vec{\kappa_3}}^* \hat{a}_{{\rm R} \vec{\kappa_4}} \Omega_{\rm M} \rangle = 0, \\
&\langle \Omega_{\rm M} | \hat{a}_{{\rm R} \vec{\kappa_1}}^* \hat{a}_{{\rm L} \vec{\kappa_2}} \hat{a}_{{\rm L} \vec{\kappa_3}}^* \hat{a}_{{\rm R} \vec{\kappa_4}} \Omega_{\rm M} \rangle = \frac{e^{- 2 \pi \omega_1/a}}{1- e^{-2 \pi \omega_1/a}} \frac{1}{1- e^{-2 \pi \omega_2/a}}\delta(\vec \kappa_1 - \vec \kappa_4) \delta(\vec \kappa_2 - \vec \kappa_3) \nonumber \\
& + \frac{e^{-  \pi \omega_1/a}}{1- e^{-2 \pi \omega_1/a}} \frac{e^{-  \pi \omega_2/a}}{1- e^{-2 \pi \omega_2/a}}\tilde \delta(\vec \kappa_1 - \vec \kappa_3) \tilde \delta(\vec \kappa_2 - \vec \kappa_4)\\
&\langle \Omega_{\rm M} | \hat{a}_{{\rm R} \vec{\kappa_1}}^* \hat{a}_{{\rm L} \vec{\kappa_2}}^* \hat{a}_{{\rm L} \vec{\kappa_3}} \hat{a}_{{\rm R} \vec{\kappa_4}} \Omega_{\rm M} \rangle = \frac{e^{- 2\pi \omega_1/a}}{1- e^{-2 \pi \omega_1/a}} \frac{e^{- 2\pi \omega_2/a}}{1- e^{-2 \pi \omega_2/a}}\delta(\vec \kappa_1 - \vec \kappa_4) \delta(\vec \kappa_2 - \vec \kappa_3) \nonumber \\
& + \frac{e^{-  \pi \omega_1/a}}{1- e^{-2 \pi \omega_1/a}} \frac{e^{-  \pi \omega_3/a}}{1- e^{-2 \pi \omega_2/a}}\tilde \delta(\vec \kappa_1 - \vec \kappa_2) \tilde \delta(\vec \kappa_3 - \vec \kappa_4),  \nonumber \\
&\langle \Omega_{\rm M} | \hat{a}_{{\rm R} \vec{\kappa_1}} \hat{a}_{{\rm L} \vec{\kappa_2}}^* \hat{a}_{{\rm L} \vec{\kappa_3}}^* \hat{a}_{{\rm R} \vec{\kappa_4}}^* \Omega_{\rm M} \rangle  = 0,\\
&\langle \Omega_{\rm M} | \hat{a}_{{\rm R} \vec{\kappa_1}}^* \hat{a}_{{\rm L} \vec{\kappa_2}} \hat{a}_{{\rm L} \vec{\kappa_3}}^* \hat{a}_{{\rm R} \vec{\kappa_4}}^* \Omega_{\rm M} \rangle  = 0,\\
&\langle \Omega_{\rm M} | \hat{a}_{{\rm R} \vec{\kappa_1}}^* \hat{a}_{{\rm L} \vec{\kappa_2}}^* \hat{a}_{{\rm L} \vec{\kappa_3}} \hat{a}_{{\rm R} \vec{\kappa_4}}^* \Omega_{\rm M} \rangle  = 0,\\
&\langle \Omega_{\rm M} | \hat{a}_{{\rm R} \vec{\kappa_1}}^* \hat{a}_{{\rm L} \vec{\kappa_2}}^* \hat{a}_{{\rm L} \vec{\kappa_3}}^* \hat{a}_{{\rm R} \vec{\kappa_4}} \Omega_{\rm M} \rangle  = 0,\\
&\langle \Omega_{\rm M} | \hat{a}_{{\rm R} \vec{\kappa_1}}^* \hat{a}_{{\rm L} \vec{\kappa_2}}^* \hat{a}_{{\rm L} \vec{\kappa_3}}^* \hat{a}_{{\rm R} \vec{\kappa_4}}^* \Omega_{\rm M} \rangle  =\frac{e^{-  \pi \omega_1/a}}{1- e^{-2 \pi \omega_1/a}} \frac{e^{-  \pi \omega_2/a}}{1- e^{-2 \pi \omega_2/a}} \tilde \delta(\vec \kappa_1 - \kappa_3) \tilde \delta(\vec \kappa_2 - \vec \kappa_4) \nonumber \\
& + \frac{e^{-  \pi \omega_1/a}}{1- e^{-2 \pi \omega_1/a}} \frac{e^{-  \pi \omega_3/a}}{1- e^{-2 \pi \omega_3/a}} \tilde \delta(\vec \kappa_1 - \kappa_2) \tilde \delta(\vec \kappa_3 - \vec \kappa_4).
\end{align}
\end{subequations} 

Following steps analogous to those in appendix \eqref{Sec:RightUpdate} we obtain eq. \eqref{DeltaL}.

\end{document}